\documentclass[11pt]{article}
\usepackage{amsmath,amssymb}
\usepackage{graphicx}
\usepackage{dsfont}

\usepackage[sc,small]{caption}
\setcaptionwidth{14cm}
\usepackage{color}

\newcommand{\mathscr}{\mathcal}

\newcommand{\be}{\begin{eqnarray}}
\newcommand{\ee}{\end{eqnarray}}

\newcommand{\J}{{\mathcal J}}
\renewcommand{\O}{{\mathcal O}}
\newcommand{\dd}{\mathrm{d}}
\newcommand{\MX}{X}
\newcommand{\idmat}{{\bf 1}}

\newcommand{\beqn}{\begin{eqnarray}}
\newcommand{\eeqn}{\end{eqnarray}}

\newcommand{\Ref}[1]{(\ref{#1})}

\begin{document}
\vspace{1cm}

\begin{center}
{\bf\LARGE
Growth Histories in Bimetric Massive Gravity \\  }

\vspace{2.5cm}

{\large
{\bf Marcus Berg$^{+ \dag}$}, 
{\bf Igor Buchberger$^{+}$, }
{\bf Jonas Enander$^{\dag}$, }\\[2mm]
{\bf Edvard M\"ortsell$^{\dag}$} ,
{\bf Stefan Sj\"ors$^{\dag}$} }
\vspace{1cm}

{\it
$^{+}$
Department of Physics, 
Karlstad University \\  651 88 Karlstad, Sweden\\
\ \\
$^{\dag}$ Oskar Klein Center, Stockholm University\\
 Albanova University Center\\ 106 91 Stockholm, Sweden\\ [5mm]
}

\end{center}
\vspace{2cm}

\begin{center}
{\bf Abstract}
\end{center}
We perform cosmological perturbation theory in Hassan-Rosen bimetric
gravity for general homogeneous and isotropic
backgrounds. In the de Sitter approximation, we obtain decoupled sets of
massless and massive scalar gravitational fluctuations. Matter perturbations 
then evolve like in Einstein gravity. We  perturb
the future de Sitter regime by the ratio of matter to dark energy, producing quasi-de Sitter space.
In this more general setting  the massive and massless fluctuations mix. We argue
that in the quasi-de Sitter regime, the growth of structure in bimetric gravity
differs from that of Einstein gravity.
 
\clearpage
 \tableofcontents

\section{Introduction}

The cosmological constant problem 
\cite{Weinberg:1988cp,Polchinski:2006gy}
is one of the most vexing problems in physics. 
The main problem is why the vacuum energy densities
of quantum field theory
seem to contribute to observable gravitational physics so much less
than simple estimates would indicate. Presumably, 
the problem would be resolved in a theory of quantum gravity. String theory contains
some quantum gravity, but it is notoriously difficult to address quantum problems in dynamical gravity with the present formulation of string theory. 
With the observation of the accelerated expansion of the universe in
1998, usually attributed to a dominant dark energy component
such as the cosmological constant (or quintessence, which has similar naturalness
problems),
the issue has been driven to a sharp point.
As reviewed in \cite{Polchinski:2006gy}, very few 
of the many proposed solutions to the problem stand a remote chance of success. 

Modified gravity is one of the well-known proposed ``solutions'' that fares particularly poorly in the evaluation of e.g. \cite{Polchinski:2006gy}, because modified gravity theories are typically {\it only} deep-infrared modifications of gravity  (where ``deep-infrared'' means very low energies, some
tiny fraction of an electron volt), which ultimately seems insufficient to solve the problem
of quantum field theory contributions from all known fields,
including for example around the electron mass of 511 keV. 
There are suggestions how modified gravity could
effectively limit how energy gravitates (a ``filter'') at a wider variety of energy scales,
as in the proposed ``degravitation'' mechanism
 \cite{Dvali:2007kt},
and the earlier discussions of screening mechanisms 
summarized in \cite{Polyakov:2009nq}. These are 
intriguing but incomplete suggestions,
in that it is not yet clear if any of these mechanisms are
actually realized in any underlying theory in which 
the range of applicability of these mechanisms could  be reliably evaluated. 

Massive gravity, a theory of gravity 
where the graviton has a mass 
(which  is typically constrained by observations to be
extremely small, perhaps $10^{-33}$ eV, see e.g.\ \cite{Sjors:2011iv}
for a list of references)
 is at face value a relatively minor and again deep-infrared modification of gravity.
The study of massive gravity was initiated in 1939 by Fierz and Pauli \cite{Fierz:1939ix}, but these theories suffered from ghost instabilities
at the  nonlinear level. In 2010 progress was made when a particular 
set of nonlinear ghost-free interactions   was found by de Rham, Gabadadze and Tolley (dRGT) in a series of papers
\cite{Burrage:2011cr,D'Amico:2011jj,deRham:2011qq,deRham:2011rn,deRham:2011by,deRham:2010kj,deRham:2010ik,deRham:2010gu} ,
following seminal earlier work in \cite{ArkaniHamed:2002sp,Creminelli:2005qk}. 
The dRGT formulation of nonlinear massive gravity requires a {\it fixed} auxiliary two-tensor 
$f_{\mu\nu}$ with no dynamics of its own. Apart
from aesthetic concerns about this, for our purposes it 
is a deficiency of this theory 
that it seems to have no homogeneous and isotropic cosmological solutions  \cite{D'Amico:2011jj,DeFelice:2012mx}.

Last year, Hassan and Rosen \cite{Hassan:2011ea,Hassan:2011zd,Hassan:2011tf}
gave dynamics to the tensor $f_{\mu\nu}$ in a 
bimetric theory, with the
nonlinear interactions between $g_{\mu\nu}$ and $f_{\mu\nu}$ imported from the de Rham-Gabadadze-Tolley massive gravity theories.
(In fact, the need for two metrics 
to realize massive gravity covariantly was already appreciated in 1976  \cite{Salam:1976as},
but there was no theory without nonlinear instabilities.)
The Hassan-Rosen bimetric theory has cosmological solutions,
as explored in \cite{Volkov:2011an,vonStrauss:2011mq,Comelli:2011zm,Volkov:2012cf}.
In the aforementioned cosmological solutions of Hassan-Rosen bimetric gravity, both 
$g_{\mu\nu}$ and $f_{\mu\nu}$ 
have equations of motion, and the background solutions
we consider are of general FLRW form for {\it both} backgrounds.

Now, bimetric gravity is a more far-reaching modification of gravity than massive gravity,
in that the new gravitationally coupled tensor field 
$f_{\mu\nu}$ has dynamics of its own. (In the formulation we will be using here, it does not couple directly to matter,
so ``bimetric'' is a little bit of a misnomer, ``gravity coupled to matter and a symmetric two-tensor'' would have been more accurate.) Of course, since the field content and interactions of bimetric theory are different from Einstein gravity, this
theory may have different quantum properties  at any given  scale and not exclusively in the deep infrared.
One way to try to understand the quantum properties of this theory would be to try to embed the theory
in string theory, but  currently it is not known how to do this. On the good side, the Hassan-Rosen bimetric theory (respectively de Rham-Gabadadze-Tolley massive gravity theory) has the kind of rigid structure that one would think could possibly descend from an underlying theory, like string theory. Symmetries constrain the interaction terms to the form $V(f^{-1}g)$, and their relative coefficients are constrained, and it is now understood how to construct these theories in various dimensions and including higher-derivative corrections \cite{Paulos:2012xe}. It would be somewhat surprising, and a shame, if this structure existed for no reason at all. 

To be clear, there is so far no clear indication that even embedding Hassan-Rosen bimetric theory in string theory would particularly help with the cosmological constant problem, but  at least the problems could be addressed in a theory that is apparently nonlinearly consistent and also fundamentally different from Einstein gravity already at the level of the low-energy effective action. 

 On the other hand, it may be easier  to rule these kinds of theories out observationally (and classically) than to properly understand their quantization, so here we pursue strategies to achieve classical observational tests. 
In this paper, we 
\begin{itemize}
\item
derive the  linearized gravitational scalar  fluctuation equations \\
for FLRW backgrounds (section \ref{sec:pert})
\item
find convenient gauge invariant variables (section \ref{sec:pert})
\item
solve the equations in the special case of a de Sitter background (section \ref{sec:dS})\\--- this is not completely 
new, see \cite{Crisostomi:2012db,Khosravi:2012rk}
\item
develop the quasi-de Sitter (qdS) approximation in bimetric gravity
 (section \ref{qdS})
\item
find solutions of the qdS fluctuation equations, both analytical and numerical (section \ref{qdS})
\item
in general, construct some necessary framework for the analysis of growth of structure
in Hassan-Rosen bimetric gravity.
\end{itemize}
Detailed observational and phenomenological analyses are left for the future.

We also mention that there has also been recent related work on multi-metric theory,
the natural generalization of this framework to coupling multiple spin-two fields nonlinearly
\cite{Khosravi:2011zi,Hinterbichler:2012cn,Hassan:2012wt}. 

Finally, there is also progress on related
theories in three dimensions \cite{deRham:2011ca}.
In fact, the ``new massive gravity'' theory
in three dimensions \cite{Bergshoeff:2009hq},
which generated some excitement in the last few years,
 is a scaling limit of the Hassan-Rosen bimetric theory \cite{Afshar:2009rg,Paulos:2012xe}.
 
 \section{Hassan-Rosen bimetric massive gravity}

The bimetric massive gravity theory found by Hassan and Rosen \cite{Hassan:2011zd}
is given by the action
\be \label{HR}
S_{\rm HR}=-\frac{M_{g}^{2}}{2}\int d^{4}x\sqrt{-g}R\left(g\right)-\frac{M_{f}^{2}}{2}\int d^{4}x\sqrt{-f}R\left(f\right)
\ee
\begin{equation}
+m^{2}M_{g}^{2}\int d^{4}x\sqrt{-g}\sum_{n=0}^{4}\beta_{n}e_{n}\left(\sqrt{g^{-1}f}\right)+\int d^{4}x\sqrt{-g}\mathcal{L}_{m}\left(g,\Phi\right).\label{eq:HRaction}
\end{equation}
and represents a natural generalization of the de Rham-Gabadadze-Tolley
massive gravity theory \cite{deRham:2010kj} to a theory with two dynamical
metrics, as discussed in the introduction. Here $\beta_{n}$ are free
parameters, which in general are the coefficients in the ``deformed
determinant'' of \cite{Hassan:2011vm}. The interaction terms $e_{n}\left(\MX\right)$
are elementary symmetric polynomials of the eigenvalues of the matrix
$\MX$, which explicitly are given by
\[
e_{0}\left(\MX\right)=1,\qquad e_{1}\left(\MX\right)=\mbox{Tr }\MX,\qquad e_{2}\left(\MX\right)=\frac{1}{2}\left(\left(\mbox{Tr }\MX\right)^{2}-\mbox{Tr }\MX^{2}\right),
\]
\begin{equation}
e_{3}\left(\MX\right)=\frac{1}{6}\left(\left(\mbox{Tr }\MX\right)^{3}-3\, \mbox{Tr }\MX\, \mbox{Tr }\MX^{2}+2\, \mbox{Tr }\MX^{3}\right),\qquad e_{4}\left(\MX\right)=\det\left(\MX\right)\label{eq:en}
\end{equation}
We have chosen to only couple $g_{\mu\nu}$ to matter,
and not $f_{\mu \nu}$, as in the original papers.
We note that this is not the only possible choice and it 
would be interesting to explore other options. 

The equations of motion are given by varying the action with respect
to $g_{\mu\nu}$ and $f_{\mu\nu}$:
\begin{equation}
R_{\mu\nu}-\frac{1}{2}g_{\mu\nu}R+\frac{m^{2}}{2}\sum_{n=0}^{3}\left(-1\right)^{n}\beta_{n}\left[g_{\mu\lambda}Y_{\left(n\right)\nu}^{\lambda}\left(\sqrt{g^{-1}f}\right)+g_{\nu\lambda}Y_{\left(n\right)\mu}^{\lambda}\left(\sqrt{g^{-1}f}\right)\right]=\frac{1}{M_{g}^{2}}T_{\mu\nu},\label{eq:geom}
\end{equation}
\begin{equation}
\bar{R}_{\mu\nu}-\frac{1}{2}f_{\mu\nu}\bar{R}+\frac{m^2}{2 M^2_\star}\sum_{n=0}^{3}\left(-1\right)^{n}\beta_{4-n}\left[f_{\mu\lambda}Y_{\left(n\right)\nu}^{\lambda}\left(\sqrt{f^{-1}g}\right)+f_{\nu\lambda}Y_{\left(n\right)\mu}^{\lambda}\left(\sqrt{f^{-1}g}\right)\right]=0,\label{eq:feom}
\end{equation}
where $\bar{R}_{\mu\nu}$ and $\bar{R}$ are the Ricci tensor and
Ricci scalar due to $f_{\mu\nu}$, and 
\begin{equation}
M_{\star}^2 \equiv\frac{M_{f}^{2}}{M_{g}^{2}}.\label{eq:mstar}
\end{equation}
The matrices $Y_{\left(n\right)\mu}^{\lambda}\left(\MX\right)$
are given by
\[
Y_{\left(0\right)}\left(\MX\right)=\idmat,\quad Y_{\left(1\right)}\left(\MX\right)=\MX-\idmat\, \mbox{Tr }\MX,
\]
\begin{equation}
Y_{\left(2\right)}\left(\MX\right)=\MX^{2}-\MX\mbox{Tr }\MX+\frac{1}{2}\, \idmat\left(\left(\mbox{Tr }\MX\right)^{2}-\mbox{Tr }\MX^{2}\right),  \label{eq:Yn} 
\end{equation}
\begin{equation}
Y_{\left(3\right)}\left(\MX\right)=\MX^{3}-\MX^{2}\, \mbox{Tr }\MX+\frac{1}{2}\MX\left(\left(\mbox{Tr }\MX\right)^{2}-\mbox{Tr }\MX^{2}\right)-\frac{1}{6}\idmat\left(\left(\mbox{Tr }\MX\right)^{3}-3\, \mbox{Tr }\MX\, \mbox{Tr }\MX^{2}+2\, \mbox{Tr }\MX^{3}\right)\; ,  \nonumber
\end{equation}
where $\idmat$ is the identity matrix.
Imposing that $T_{\mu\nu}$ is covariantly conserved, then  from eq.~(\ref{eq:geom}),
the Bianchi constraint gives
\begin{equation}
\nabla^{\mu}\sum_{n=0}^{3}\left(-1\right)^{n}\beta_{n}\left[g_{\mu\lambda}Y_{\left(n\right)\nu}^{\lambda}\left(\sqrt{g^{-1}f}\right)+g_{\nu\lambda}Y_{\left(n\right)\mu}^{\lambda}\left(\sqrt{g^{-1}f}\right)\right]=0.\label{eq:bianchi}
\end{equation}
It can be shown that the corresponding Bianchi constraint given from
(\ref{eq:feom}) is equivalent with (\ref{eq:bianchi}).

Finally, by performing the constant rescaling
\begin{equation}
f_{\mu\nu}\rightarrow\frac{M_{g}^{2}}{M_{f}^{2}}f_{\mu\nu},\qquad\beta_{n}\rightarrow\left(\frac{M_{f}}{M_{g}}\right)^{n}\beta_{n},\label{eq:fandbetarescale}
\end{equation}
we set $M_{\star}^{2}$  to unity. In other words,
$M_{\star}^{2}$ was  a redundancy that we do not consider a separate free parameter.

\section{Review of background equations}

Cosmological solutions of the Hassan-Rosen bimetric theory were studied
in \cite{Volkov:2011an,vonStrauss:2011mq,Comelli:2011zm,Volkov:2012cf}. 
We have a separate metric ansatz for each of $g_{\mu\nu}$ and $f_{\mu\nu}$ (specialized to the case
of flat spatial sections):
\begin{equation}
ds_{g}^{2}=-dt^{2}+a^{2}\left(t\right)d\bar{x}^{2},\label{eq:gansatz}
\end{equation}
\begin{equation}
ds_{f}^{2}=-X^{2}\left(t\right)dt^{2}+Y^{2}\left(t\right)d\bar{x}^{2}.\label{eq:fansatz}
\end{equation}
The Bianchi constraint given in eq.~(\ref{eq:bianchi}) gives
\begin{equation}
\frac{3m^{2}}{a}\left(\beta_{1}+2\frac{Y}{a}\beta_{2}+\frac{Y^{2}}{a^{2}}\beta_{3}\right)\left(\dot{Y}-\dot{a}X\right)=0,\label{eq:bianchicosmological}
\end{equation}
where overdots denote differentiation with respect to $t$. 
There are two options for solving the Bianchi identity,
which we refer to as Case A and Case B. In Case A, which we will not use,
\begin{equation}
\beta_{1}+2\frac{Y}{a}\beta_{2}+\frac{Y^{2}}{a^{2}}\beta_{3}=0,\label{eq:bianchi1}
\qquad \mbox{(not used)}
\end{equation}
which gives a cosmological solution
that is degenerate with GR, for which the fluctuation equations reduces to identical
equations to those of GR (as noted in \cite{Comelli:2011zm}).
Thus, we  focus on Case B, which is
\begin{equation}
X=\frac{\dot{Y}}{\dot{a}}.\label{eq:bianchi2}
\end{equation}
The Friedmann equations derived from eq.~(\ref{eq:geom}) and eq.~(\ref{eq:feom}),
together with the Bianchi identity, are
\begin{equation}
-3\left(\frac{\dot{a}}{a}\right)^{2}+m^{2}\left(\beta_{0}+3\beta_{1}\frac{Y}{a}+3\beta_{2}\frac{Y^{2}}{a^{2}}+\beta_{3}\frac{Y^{3}}{a^{3}}\right)=\frac{1}{M_{g}^{2}}T_{0}^{0},\label{eq:gfriedmann}
\end{equation}
\begin{equation}
-3\left(\frac{\dot{a}}{Y}\right)^{2}+m^{2}\left(\beta_{1}+3\beta_{3}\frac{a}{Y}+3\beta_{2}\frac{a^{2}}{Y^{2}}+\beta_{3}\frac{a^{3}}{Y^{3}}\right)=0.\label{eq:ffriedmann}
\end{equation}
The acceleration equations can be shown to follow from the two Friedmann equations when using the Bianchi constraint. 

In this paper we will only consider the simplest class of solutions,
corresponding to $\beta_{1}=\beta_{3}=0$, as discussed in \cite{vonStrauss:2011mq}.
Combining the two Friedmann equations then gives
\begin{equation}
H^{2}=\frac{\beta_{4}}{\beta_{4}-3\beta_{2}}\frac{\rho}{3M_{g}^{2}}+m^{2}\frac{\beta_{0}\beta_{4}-9\beta_{2}^{2}}{3\left(\beta_{4}-3\beta_{2}\right)},\label{eq:Hfriedmann1}
\end{equation}
\begin{equation}
\frac{Y^{2}}{a^{2}}=\frac{\rho}{m^{2}M_{g}^{2}\left(\beta_{4}-3\beta_{2}\right)}+\frac{\beta_{0}-3\beta_{2}}{\beta_{4}-3\beta_{2}},\label{eq:Y/a1}
\end{equation}
where $H={\dot{a}}/{a} $ and $\rho=-T_{0}^{0}$
corresponds to the pressureless matter density.

We now define four effective parameters, to be used in our fluctuation
analysis, according to
\[
H_{\rm dS}^{2}\equiv m^{2}\frac{\beta_{0}\beta_{4}-9\beta_{2}^{2}}{3\left(\beta_{4}-3\beta_{2}\right)},\quad M_{P}^{2}\equiv M_{g}^{2}\frac{\beta_{4}-3\beta_{2}}{\beta_{4}}
\]
\begin{equation}
M^{2}\equiv2m^{2}\left(1+c^{2}\right)\beta_{2},\quad c^{2}\equiv\frac{\beta_{0}-3\beta_{2}}{\beta_{4}-3\beta_{2}}.\label{eq:effectiveparameters}
\end{equation}
The importance of these particular combinations 
of parameters in the action  (\ref{eq:HRaction}) is as follows. 
First observe in the action that $\beta_{0}$ 
can be thought of as setting 
the usual $g$ cosmological constant,
that $\beta_{4}$ likewise can be thought of as setting 
the $f$ cosmological constant, but 
that the effective ``observable'' cosmological constant
that actually appears in 
\Ref{eq:Hfriedmann1}  is a combination of 
$\beta_0$,  $\beta_{2}$ and
$\beta_{4}$. 
In the  $\rho\rightarrow0$
limit of \Ref{eq:Hfriedmann1},
there is a de Sitter solution,
and its Hubble constant $H_{\rm dS}^{2}$ is then 
related to the effective cosmological
constant induced by the interaction potential,
that is in turned fixed by the $\beta_n$ parameters.
If we also consider  $\rho\rightarrow0$
in \Ref{eq:Y/a1}, we see that also $f_{\mu\nu}$ will have a de Sitter solution
with possibly different overall normalization,
and the parameter
$c$ is the proportionality constant between $g_{\mu\nu}$ and $f_{\mu\nu}$
in the de Sitter spacetime. Further, $M^{2}$ is the mass of the spin-2 helicity
modes when linearizing $g_{\mu\nu}$ and $f_{\mu\nu}$ around such proportional
background metrics (see appendix \ref{app:tensor}). Finally, $M_{P}^{2}$
is the effective gravitational coupling constant for $\rho$ in the cosmological
framework (note that this will not be the coupling constant for the
fluctuations, nor does it necessarily describe the coupling 
in local solutions).

In terms of these parameters, the $H$ equation can  be written
\begin{equation}
H^{2}=\frac{\rho}{3M_{P}^{2}}+H_{\rm dS}^{2} \; . \label{eq:Hfriedmann2}
\end{equation}
As usual,  the continuity equation for equation of state $p=w\rho$ 
with constant equation of state parameter $w$ reads
\be  \label{cont}
{d \ln \rho \over d \ln a} = -3(1+w)
\ee
with solution $\rho = \rho_0 a^{-3(1+w)}$.
Pressureless matter ($w=0$) evolves as $a^{-3}$, and normalizing the scale factor at the present time to $a_0=1$, we can rewrite  eqs.~(\ref{eq:Hfriedmann2}) 
and \Ref{eq:Y/a1} as
\be
  \left(\frac{H}{H_0}\right)^2&=&\frac{1}{a^3}\left[1-\left(\frac{H_{\rm dS}}{H_0}\right)^2\right]+\left(\frac{H_{\rm dS}}{H_0}\right)^2  \label{Fried3} \\ [2mm]
\frac{Y^{2}}{a^{2}}&=&c^{2}\frac{2\left(1+c^{2}\right)H^{2}-M^{2}}{2\left(1+c^{2}\right)H_{\rm dS}^{2}-M^{2}}.\label{eq:Y/a2}
\ee   
where we also used the relation
\begin{equation}
M_{P}^{2}=M_{g}^{2}\left(1+c^{2}\right)\frac{M^{2}-2H_{\rm dS}^{2}}{M^{2}-2\left(1+c^{2}\right)H_{\rm dS}^{2}}
\label{eq:Mpl} \; . 
\end{equation}

Equations \Ref{Fried3} and \Ref{eq:Y/a2} are our final forms of the background equations,
expressed entirely in terms of the parameters \Ref{eq:effectiveparameters}.

In terms of background cosmology, since the form \Ref{Fried3} is equivalent to $\Lambda$CDM, the only relevant parameter is $H_{\rm dS}^2/H_0^2$, which can be constrained by observational data to be close to $0.7$, by relating it to the usual ratios to critical densities:
\be
\Omega_{\Lambda} = {H_{\rm dS}^2 \over H_0^2} \;  , \quad \Omega_m = 1-{H_{\rm dS}^2 \over H_0^2}  \; . 
\label{Omegas}
\ee   
It is therefore only the specific 
combination of $\beta_n$ given by the definition of $H_{\rm dS}$ that is constrained by the expansion history of the universe, leaving $M_{P}^2$, $M^2$ and $c^2$ as unconstrained parameters, possibly to be constrained 
by structure formation data, but there are some further restrictions, as we shall see.

Finally, a comment on the range of these parameters,
in particular of the mass parameter $M$.
Since $H^{2}\geq H_{\rm dS}^{2}$ always (see e.g. fig.\ \ref{fig:H}), we note that 
if $2\left(1+c^{2}\right)H_{\rm dS}^{2}< M^{2}$
and at some time it happens that
 $2\left(1+c^{2}\right)H^{2}(t)\geq M^{2}$ (which can occur in the early universe), 
 then from  \Ref{eq:Y/a2}
the $f$ scale factor $Y$ will be {\it imaginary},  
which we consider unphysical.\footnote{It might be interesting
to explore this branch of solutions, for example by
picking a sufficiently large $M$ that  moves this region to the very early universe
where the current model is in any case  not applicable. We will not consider such
models in this paper.}
Therefore, we demand that $M^{2}\leq2\left(1+c^{2}\right)H_{\rm dS}^{2}$. But then, we
see that $M_{P}^{2}$ will be {\it negative} if also $
M^{2} > 2H_{\rm dS}^{2}$. Negative values of 
 $M_{P}^2$ would 
be unphysical, since the matter density would then need to be negative in order to have expanding background solutions originating in a hot and dense 
state (as demanded by observations of the cosmic microwave backround).  

To summarize, if we demand that $Y$ should be real and $M_{P}^{2}$ positive,
we require
 \be
M^{2} < 2H_{\rm dS}^{2}   \; . 
\ee
 In these bimetric models, we thus need to violate the Higuchi bound $M^{2}>2H_{\rm dS}^{2}$ \cite{Higuchi:1986py,Deser:2001wx}
already at the level of the background. We will comment more on this later,
and see also fig.\ \ref{fig:M}. 

\section{Perturbations}
\label{sec:pert}
We will first consider a general background for $g_{\mu\nu}$ and $f_{\mu\nu}$. 
In this paper we will only consider scalar gravitational perturbations,
except for a brief review of tensor perturbations in appendix \ref{app:tensor}. 
For the scalar perturbations, following Weinberg \cite{Weinberg:2008zzc}
 we make the ansatz
\be
\label{pert}
ds_{g}^{2}&=&-\left(1+E_{g}\right)dt^{2}+2a\partial_{i}F_{g}dx^{i}dt+a^{2}\big(\left(1+A_{g}\right)\delta_{ij}+\partial_{i}\partial_{j}B_{g}\big)dx^{i}dx^{j} \\
ds_{f}^{2}&=&-X^{2}\left(1+E_{f}\right)dt^{2}+2XY\partial_{i}F_{f}dx^{i}dt+Y^{2}\big(\left(1+A_{f}\right)\delta_{ij}+\partial_{i}\partial_{j}B_{f}\big)dx^{i}dx^{j}
\ee
so our set of eight (non-gauge-invariant) scalar gravitational fluctuations
is $\{ E_g, F_g, A_g, B_g\}$ and $\{ E_f, F_f, A_f, B_f \}$. 
Note that at this point, the ansatz is completely
symmetric between the $g$ and $f$ metrics, as far as the perturbations are concerned. 

\subsection{Gauge invariant variables}
We form gauge invariant combinations of the perturbations in \eqref{pert}.
As usual there is no uniqueness in the choice of gauge invariant variables
(any combination of gauge invariant variables is gauge invariant)
but we find the following variables convenient:
\be
\begin{array}{rclrcl}
\Psi_{g}&=&-{1 \over 2}{A_{g}}+\frac{H}{2}\left[a^{2}\dot{B}_{g}-2aF_{g}\right]&\hspace{1cm} \Psi_{f}&=&-\frac{1}{2}A_{f}+\frac{K}{2X}\left[\frac{Y{}^{2}}{X}\dot{B}_{f}-2YF_{f}\right]    \\
\Phi_{g}&=&{1 \over 2}{E_{g}}-\frac{1}{2}\left[a^{2}\dot{B}_{g}-2aF_{g}\right]^{\cdot}
& \Phi_{f}&=&{1 \over 2}{E_{f}}-\frac{1}{2X}\left[\frac{Y{}^{2}}{X}\dot{B}_{f}-2YF_{f}\right]^{\cdot}     \\
\mathcal{B}&=&\frac{1}{2}\left(B_{f}-B_{g}\right)& \mathcal{F}&=&F_{f}-\frac{aX}{Y}F_{g}+\frac{1}{2}\left[\frac{Xa^{2}}{Y}\dot{B}_{g}-\frac{Y}{X}\dot{B}_{f}\right]  
\end{array}     \label{var}
\ee
with the definitions
\be
H\equiv {\dot{a}}/{a} \; , \qquad K\equiv {\dot{Y}}/{Y} \; ,
\ee
so $K$ is the Hubble function for the $f$ metric. 
The ansatz \Ref{var} is roughly speaking ``as symmetric as possible'' between $g$ and $f$, 
but complete symmetry is unattainable
as the backgrounds are  generically different.

\subsection{Equations of motion: general background}
\label{eom}
Using the gauge invariant variables in the previous section,
the equations of motion for the scalar perturbations in the $g$ sector  become
\be
-\frac{1}{a^{2}}\nabla^{2}\Psi_{g}+3H\left(H\Phi_{g}+\dot{\Psi}_{g}\right)+\frac{m^{2}YP}{2a^{3}}\left[3\left(-\Psi_{f}+\Psi_{g}-\frac{YK}{X}\mathcal{F}\right)+\nabla^{2}\mathcal{B}\right]=\frac{1}{2M_{g}^{2}}\delta T_{0}^{0}
\label{general1}
\ee
\be
-\partial_{i}\left(\dot{\Psi}_{g}+H\Phi_{g}\right)+\frac{m^{2}YXP}{2a\left(aX+Y\right)}\partial_{i}\left(\mathcal{F}+\frac{Y}{X}\dot{\mathcal{B}}\right)=\frac{\delta T_{i}^{0}}{2M_{g}^{2}}
\ee
\be  
&&\hspace{-1cm} \ddot{\Psi}_{g}+H\dot{\Phi}_{g}+3H\left(H\Phi_{g}+\dot{\Psi}_{g}\right)+2\dot{H}\Phi_{g}+\frac{1}{2a^{2}}\left(\partial_{j}^{2}+\partial_{k}^{2}\right)\left(\Phi_{g}-\Psi_{g}\right)+\\
&&+\frac{m^{2}}{2a^{2}}\left\{P \left[X\left(\Phi_{f}-\Phi_{g}\right)-(Y\mathcal{F})^{\bullet}\right]+YQ\left[2\left(-\Psi_{f}+\Psi_{g}-\frac{YK}{X}\mathcal{F}\right)+\left(\partial_{j}^{2}+\partial_{k}^{2}\right)\mathcal{B}\right]\right\} =\frac{1}{2M_{g}^{2}}\delta T_{i}^{i} \nonumber
\ee
\be
-\frac{1}{2a^{2}}\partial^i\partial_j\left(\Phi_{g}-\Psi_{g}\right)-\frac{m^{2}YQ}{2a^{2}}\partial^i\partial_j\mathcal{B}=\frac{1}{2M_{g}^{2}}\delta T_{j}^{i}
\ee
where $j,k$ are not equal to $i$.
In the $f$-sector we have
\be
-\frac{1}{Y^{2}}\nabla^{2}\Psi_{f}+3\frac{1}{X^{2}}K\left(K\Phi_{f}+\dot{\Psi}_{f}\right)-\frac{m^{2}a}{2Y^{3}}P\left[3\left(-\Psi_{f}+\Psi_{g}-\frac{YH}{X}\mathcal{F}\right)+\nabla^{2}\mathcal{B}\right]=0
\ee
\be
-\frac{1}{X^{2}}\partial_{i}\left(\dot{\Psi}_{f}+K\Phi_{f}\right)-\frac{m^{2}P}{2X^{2}\left(aX+Y\right)}\partial_{i}\left(\mathcal{F}+\frac{a^{2}X}{Y}\dot{\mathcal{B}}\right)=0
\ee
\be
&&\hspace{-1cm}\frac{1}{X^{2}}\ddot{\Psi}_{f}-\frac{\dot{X}}{X^{3}}\dot{\Psi}_{f}+\frac{1}{X^{2}}K\dot{\Phi}_{f}+3\frac{K}{X^{2}}\left(K\Phi_{f}+\dot{\Psi}_{f}\right)+\frac{2}{X}\left(\frac{K}{X}\right)^{\!\bullet}\Phi_{f}+\frac{1}{2Y^{2}}\left(\partial_{j}^{2}+\partial_{k}^{2}\right)\left(\Phi_{f}-\Psi_{f}\right)+\\
&&-\frac{m^{2}}{2XY^{2}}\left\{ P\left[\Phi_{f}-\Phi_{g}-\left(\frac{Y}{X}\mathcal{F}\right)^{\!\bullet}\right]+aQ\left[2\left(-\Psi_{f}+\Psi_{g}-\frac{YH}{X}\mathcal{F}\right)+\left(\partial_{j}^{2}+\partial_{k}^{2}\right)\mathcal{B}\right]\right\}=0\nonumber \ee
\be
-\frac{1}{2Y^{2}}\partial^i\partial_j\left(\Phi_{f}-\Psi_{f}\right)+\frac{m^{2}aQ}{2XY^{2}}\partial^i\partial_j\mathcal{B}=0
\label{general8}
\ee
with the definitions
\be  \label{defs}
P\equiv\left(\beta_{1}a^{2}+2\beta_{2}aY+\beta_{3}Y^{2}\right)\;,\quad Q\equiv\left[a\beta_{1}+\beta_{2}\left(aX+Y\right)+\beta_{3}XY\right]\; .
\ee
At this point we
recall that although the separate Einstein-Hilbert actions 
for $g$ and $f$ in eq.\  \Ref{HR} 
are of course invariant under separate diffeomorphisms
of the two metrics,
the mass terms are only invariant under diagonal diffeomorphisms
that preserve $g^{-1}f$. 
Thus, the $\Psi_g$, $\Phi_g$ and
$\Phi_f$, $\Psi_f$ can only appear as {\it differences}
in the mass terms, and we see this manifestly in the equations.

\subsection{Massless  limit}
If we were to turn off the interaction potential
in the action, we might expect to  find two decoupled sets of fluctuations. In fact we observe that
if $\beta_1=\beta_2=\beta_3=0$  then the two combinations
 $P$ and $Q=0$ both vanish, so 
 ${\mathcal F}$ and ${\mathcal B}$ drop out of the equations entirely,
 and the $g$ fluctuations and $f$ fluctuations constitute decoupled sectors. 
 
 This is a simple  observation purely in terms of the fluctuation equations. 
 However, whether the fluctuations truly represent decoupled physics is a subtle issue.
 For example, the condition \ \Ref{eq:bianchi2} 
 from the Bianchi identity relates
 the $g$ and $f$ background solutions for arbitrarily small $M$,
 but there is a priori no reason to impose this condition  in the strictly massless theory. (In the language of that section, one can revert to Case A, in which case \Ref{eq:bianchi2} need not be imposed.)
  But if the condition imposed on the background differs between
  $M\rightarrow 0$ and $M=0$, there is a potential 
  ``cosmological vDVZ discontinuity''
 \cite{vanDam:1970vg,Zakharov:1970cc},   i.e. the  $M
 \rightarrow 0$ and $M=0$ theories
  could potentially be different no matter how small $M$ is taken in the limit.  
  Of course, there could still be a Vainshtein mechanism 
  \cite{Vainshtein:1972sx} that resolves the discontinuity in the 
  nonlinear regime, but this would not be evident in our linear approximation.
In  figure  \ref{fig:delta} below, we see some hint of a discontituity,
but it is somewhat subtle here as we have several parameters to play with. 
We will not resolve the issue
of the existence of a smooth limit here, but see also the recent interesting discussions by
\cite{Baccetti:2012bk,Paulos:2012xe}.

Now we consider more special
backgrounds, first a
two-component fluid solution
that we will refer to as the ``exact solution'', then de Sitter and then quasi-de Sitter. 

\section{Exact solution}
It is well-known that in the approximation
of a two-component fluid of pressureless matter (dust) with  equation of 
state $p=0$ ($w=0$) and cosmological constant with  equation of state $p=-\rho$
($w=-1$),
the combination of equations \Ref{Fried3} and \Ref{cont}
admits the exact solution
\be  \label{exact}
a(t) = c_1 \sinh\left({3 \over 2}H_{\rm dS} t\right)^{2/3}
\ee
where the constants $c_1$ and $H_{\rm dS}$ are
\be
c_1 = \left({\Omega_m \over \Omega_{\Lambda}}\right)^{1/3} =
\left(1-\Omega_{\Lambda} \over \Omega_{\Lambda}\right)^{1/3} \; , \quad
H_{\rm dS} = \sqrt{\Omega_{\Lambda}}H_0  \; .
\ee
%From \Ref{Upsilon} we find an expression for $Y$ as a sum of powers of $a(t)$:
%\be
%Y(t)^2 = c_3 a(t)^2 + c_4 a(t)^{-1}
%\ee
%where
%\be
%c_3= {3m^2+6 (1-\Omega^{\rm eff}_{\Lambda})  \over 2\beta_4 m^2} \; ,
%\quad c_4 = {3 \Omega_{\Lambda}^{\rm eff} \over m^2 \beta_4} \; .
%\ee
We can write $\dot{a}=Ha$ in \Ref{eq:gfriedmann} to
express $Y(t)$ in terms of $a(t)$:
\be 
Y(t) = {\sqrt{3}  \over \sqrt{\beta_4}m}  \sqrt{H^2(t)-\beta_2 m^2}
\cdot  a(t) \; . 
\ee
and then using the effective parameters
\Ref{eq:effectiveparameters} we obtain for the scale factor
of the $f$ metric:
\be
Y=\sqrt{\frac{2\left(1+c^{2}\right)H_{\rm dS}^{2}\coth^{2}\left(\frac{3}{2}H_{\rm dS}t\right)-M^{2}}{2\left(1+c^{2}\right)H_{\rm dS}^{2}-M^{2}}} \cdot  c \cdot c_{1}\sinh\left(\frac{3}{2}H_{\rm dS}t\right)^{2/3}
 \label{Y2}
 \ee
Ideally one would now simply use these
 background scale factors in the fluctuation equations and solve them numerically,
 which would lead to a model for growth of structure
 in bimetric theory at any time $t$. 
Unfortunately, we have not been able to complete this program, and instead we will
focus on special cases and simplifying approximations. 

The simplest special case is that $H=$ constant  as in pure de Sitter space, 
then \Ref{Y2} tells us that $Y(t) \propto a(t)$, with the constant of proportionality
given by
\be  \label{constprop}
c \; \equiv \; 
{Y_{\rm dS}(t) \over a_{\rm dS}(t)} \; . 
\ee
We will in general not limit ourselves to pure de Sitter space, 
but it provides a useful starting point. Any departure 
of the $g_{\mu\nu}$ metric from pure de Sitter space breaks the proportionality
between the $g_{\mu\nu}$ scale factor and the $f_{\mu\nu}$ scale factor.

\section{Pure de Sitter space}
\label{sec:dS}
Matter dilutes
away as the universe expands, and the universe approaches a  de Sitter (pure dark energy) solution in the future. To provide some feeling for the numbers, if the evolution would proceed
according to GR, then it will take around 10 Gyr after present for the exact FLRW scale factor $a(t)$ to 
agree with the de Sitter scale factor $a_{\rm dS}(t)$ to within 1\% accuracy. 

For large time, the exact solutions in the previous sections reduce to 
the approximate solutions $a(t)\rightarrow a_{\rm dS}(t)$, $Y(t)\rightarrow Y_{\rm dS}(t)$ where
\be  \label{dSsol}
a_{\rm dS}(t) &=& c_2
\exp\left(H_{\rm dS} t \right) \\
Y_{\rm dS}(t) &=& c \cdot c_2  \exp\left(H_{\rm dS} t \right)
\ee
where
\be
c_2 &=&  \left( {1 - \Omega_{\Lambda} \over 4 \Omega_{\Lambda}}\right)^{1/3} 
\ee
and  $c$ is the proportionality constant from \Ref{Y2}.
Although these are of course exact de Sitter
solutions in their own right, it is useful to consider them
as limits of the exact solution for normalization purposes. 
In particular, since there is no Big Bang in pure de Sitter,
there would have been no way to normalize the scale factor. 

\subsection{Gauge invariant variables in dS}

The general gauge invariant variables of \Ref{var} have the following dS limits:
\be   
\begin{array}{rclrcl}
\Psi_{g}&=&-{1 \over 2}{A_{g}}+\frac{H_{\rm dS}}{2}\left[a^{2}\dot{B}_{g}-2aF_{g}\right]&\hspace{1cm} \Psi_{f}&=&-{1 \over 2}{A_{f}}+\frac{H_{\rm dS}}{2}\left[a^{2}\dot{B}_{f}-2aF_{f}\right]   \\
\Phi_{g}&=& {1 \over 2}{E_{g}}-\frac{1}{2}\left[a^{2}\dot{B}_{g}-2aF_{g}\right]^{\cdot}& \Phi_{f}&=&{1 \over 2}{E_{f}}-\frac{1}{2}\left[a^{2}\dot{B}_{f}-2aF_{f}\right]^{\cdot} \\
\mathcal{B}&=&\frac{1}{2}\left(B_{f}-B_{g}\right)& \mathcal{F}&=&F_{f}-F_{g}+\frac{a}{2}\left[\dot{B}_{g}-\dot{B}_{f}\right]
\end{array}
\label{vardS}
\ee
Defining the linear combinations of fields\footnote{Had we not set $M_{\star}=1$ by rescaling,
it would also enter in these combinations.}
\be
\Phi_{+}=\Phi_{g}+c^{2}\Phi_{f}\,\,\,\,\,\,\,\,\,\,\,\,\,\,\,\,\,\,\,\,\Phi_{-}=\Phi_{g}-\Phi_{f}
\ee
\be
\Psi_{+}=\Psi_{g}+c^{2}\Psi_{f}\,\,\,\,\,\,\,\,\,\,\,\,\,\,\,\,\,\,\,\,\Psi_{-}=\Psi_{g}-\Psi_{f}
\ee
we are able 
to separate the scalar gravitational fluctuation equations \Ref{general1}-\Ref{general8}
into a system of massless equations for $\Phi_+$, $\Psi_+$ 
and  massive
equations for $\Phi_-$, $\Psi_-$.
The 
${\mathcal F}$ and ${\mathcal B}$ fields appear only in
the massive field equations.
The matter perturbations will appear in both sectors.

\subsection{Perturbations in dS}

The equations of motion reduce as follows.
In the massless sector:
\be  \label{massless1}
-\frac{1}{a^{2}}\nabla^{2}\Psi_{+}+3H_{\rm dS}\left(H\Phi_{+}+\dot{\Psi}_{+}\right)=\frac{\delta T_{0}^{0}}{2M_{g}^{2}}
\ee
\be   \label{massless2}
-\partial_{i}\left(\dot{\Psi}_{+}+H_{\rm dS}\Phi_{+}\right)=\frac{\delta T_{i}^{0}}{2M_{g}^{2}}
\ee
\be   \label{massless3}
\ddot{\Psi}_{+}+H\dot{\Phi}_{+}+3H_{\rm dS}\dot{\Psi}_{+}+3H_{\rm dS}^{2}\Phi_{+}+\frac{1}{2a^{2}}\left(\partial_{j}^{2}+\partial_{k}^{2}\right)\left(\Phi_{+}-\Psi_{+}\right)=\frac{\delta T_{i}^{i}}{2M_{g}^{2}}
\ee
\be    \label{massless4}
-\frac{1}{2a^{2}}\partial^{i}\partial_{j}\left(\Phi_{+}-\Psi_{+}\right)=\frac{\delta T_{j}^{i}}{2M_{g}^{2}}  \; .
\ee
We immediately note that these equations are of exactly the same form
as the analogous equations for perturbations in Einstein gravity. 
In the massive sector:
\be   \label{massive1}
-\frac{1}{a^{2}}\nabla^{2}\Psi_{-}+3H_{\rm dS}\left(H\Phi_{-}+\dot{\Psi}_{-}\right)+\frac{m^{2}P}{2a^{2}}\left(\frac{1+c^{2}}{c}\right)\left(3\Psi_{-}-3aH_{\rm dS}\mathcal{F}+\nabla^{2}\mathcal{B}\right)=\frac{\delta T_{0}^{0}}{2M_{g}^{2}}
\ee
\be   \label{massive2}
-\partial_{i}\left(\dot{\Psi}_{-}+H_{\rm dS}\Phi_{-}\right)+\frac{m^{2}P}{4a}\left(\frac{1+c^{2}}{c}\right)\partial_{i}\left(\mathcal{F}+a\dot{\mathcal{B}}\right)=\frac{\delta T_{i}^{0}}{2M_{g}^{2}}
\ee
\be   \label{massive3}
&&  \hspace{-1cm}\ddot{\Psi}_{-}+H_{\rm dS}\left(3\dot{\Psi}_{-}+\dot{\Phi}_{-}\right)+3H_{\rm dS}^{2}\Phi_{-}+\frac{1}{2a^{2}}\left(\partial_{j}^{2}+\partial_{k}^{2}\right)\left(\Phi_{-}-\Psi_{-}\right)+   \\
 &&
+\frac{m^{2}P}{2a^{2}}\left(\frac{1+c^{2}}{c}\right)\Biggl\{\left[-\Phi_{-}-\left(a\mathcal{F}\right)^{\bullet}\right]+\left[2\Psi_{-}-2H_{\rm dS}a\mathcal{F}+\left(\partial_{j}^{2}+\partial_{k}^{2}\right)\mathcal{B}\right]\Biggr\}=\frac{\delta T_{i}^{i}}{2M_{g}^{2}}   \nonumber
\ee
\be   \label{massive4}
-\frac{1}{2a^{2}}\partial^{i}\partial_{j}\left(\Phi_{-}-\Psi_{-}\right)-\frac{m^{2}P}{2a^{2}}\left(\frac{1+c^{2}}{c}\right)\partial_{i}\partial_{j}\mathcal{B}=\frac{\delta T_{j}^{i}}{2M_{g}^{2}}
\ee
where $P$ and $Q$ are defined in \Ref{defs}. 
The earlier statement 
that only {\it differences}  of the
$\Phi$ and $\Psi$ fields can appear in the mass terms
now translates into the statement that mass terms only appear 
for the $\Phi_-$ and $\Psi_-$ fields. Thus it
must be that the $\Phi_+$ and $\Psi_+$  fields 
are massless, which is clear above.
This was also observed in \cite{Crisostomi:2012db}.

\subsection{dS solutions: Massless sector}

%The solution for $\Psi_{+,\rm dS}$, when $\delta T_{j}^{i}=0$, is 
%\begin{equation}
%\Psi_{+,dS}=\Phi_{+,dS}=\frac{C_{1}}{a^{}}+\frac{C_{2}}{a^{3}}.\label{eq:Psi+}
%\end{equation}
%\igor{I don't understand $a^3$, $a^5$; the solution is  $e^{-\sqrt{\Omega_{\Lambda}} H_0 t}, e^{-3\sqrt{\Omega_{\Lambda}} H_0 t}$ so I would say  $a$, $a^3$, different conventions? or my mistake? }

%In de Sitter there is thus only decaying modes, which means that all
%{}``classical hair'' will be negligible in less than a Hubble time
%\jon{references}. 
We will consider only pressureless matter (dust), both for background and for fluctuations.
This means that $\delta T^i_j=0$. However we will only literally use this condition in this section,
since we will have sources generating effective pressure and
anisotropic stress in the next section. 
Also, we will spatially Fourier transform the perturbations, i.e.\ 
write Fourier modes with spatial wave number $k=|{\mathbf k}|$.
For practical purposes these wave numbers will correspond to wavelengths
below the horizon scale, so $k>H_{\rm dS}$. 

With  this understanding,
the  massless (``plus'') gravitational potentials in dS are
%\begin{equation}
%\delta\rho_{dS}=2M_{g}^{2}\left[\frac{\nabla^{2}}{a^{2}}\left(\frac{C_{1}}{a^{3}}+\frac{C_{2}}{a^{5}}\right)+6H_{\rm dS}^{2}\left(\frac{C_{1}}{a^{3}}+\frac{2C_{2}}{a^{5}}\right)\right],\label{eq:deltarho}
%\end{equation}
%\begin{equation}
%\delta u_{dS}=4M_{g}^{2}H_{\rm dS}\left(\frac{C_{1}}{a^{3}}+\frac{2C_{2}}{a^{5}}\right).\label{eq:deltau}
%\end{equation}
\be  \label{Psiplus}
\Psi_{+,\rm dS}= \Phi_{+,\rm dS} = C_1 e^{-H_{\rm dS}t} + C_2 e^{-3H_{\rm dS}t}
\ee
where as before  $H_{\rm dS}=\sqrt{\Omega_{\Lambda}} H_0$.
The matter perturbations are given from the $00$ and $0i$ component
of the equations of motion:
\be  \label{rhodS}
{\delta\rho_{\rm dS} \over M_g^2} &=& 
-12 H_{\rm dS}^2 C_2 e^{-3H_{\rm dS}t} - {2\over c_5^2} k^2 (C_1 e^{-3H_{\rm dS}t}
+C_2 e^{-5H_{\rm dS}t}) \\
\label{udS}
{\delta u_{\rm dS} \over M_g^2 }&=& 4 C_2 H_{\rm dS} e^{-3 H_{\rm dS}t} \; . 
\ee
Again these are identical with the corresponding GR solutions.
As in GR, there are two integration constants $C_1$ and $C_2$,
and the remaining fields are determined
from these. 

\subsection{dS solutions: Massive sector}
\label{massivewave}

The massive (``minus") sector can be manipulated to yield a single wave equation
from which the other fields are determined.
Solving for $\mathcal F$ and $\mathcal B$ 
gives
\begin{equation}
{\mathcal{F}}=\frac{2M_{g}^{-2}\delta u+4H\Phi_{-}+4\dot{\Psi}_{-}}{aM^{2}}
-a \dot{\mathcal{B}} \label{eq:Ftilde}
\end{equation}
\begin{equation}
\mathcal{B}=\frac{\Psi_{-}-\Phi_{-}-M_{g}^{-2}\chi a^{2}}{a^{2}M^{2}}\label{eq:B}
\end{equation}
where 
the anisotropic stress $\chi$ is defined through $\delta T_{j}^{i}=\partial_{i}\partial_{j}\chi$.
%The $00$ and $ii$ equations of motion then become
%\begin{equation}
%-3\left[H^{2}-M^{2}\right]\Psi_{-}-\frac{\nabla^{2}\left(\Phi_{-}+\Psi_{-}\right)}{2a^{2}}-\frac{3}{2}H\left(\dot{\Psi}_{-}+\dot{\Phi}_{-}\right)=\frac{\delta T_{0}^{0}+6H\delta u+\nabla^{2}\chi+3Ha^{2}\dot{\chi}}{2M_{g}^{2}}\label{eq:00eq}
%\end{equation}
%\[
%\left[H^{2}-M^{2}\right]\left(\Phi_{-}-2\Psi_{-}\right)-H\left(\dot{\Phi}_{-}+\dot{\Psi}_{-}\right)-\frac{1}{2}\left(\ddot{\Phi}_{-}+\ddot{\Psi}_{-}\right)=
%\]
%\begin{equation}
%=\frac{\frac{1}{3}\delta T_{k}^{k}+\frac{2}{3}\nabla^{2}\chi+4Ha^{2}\dot{\chi}+a^{2}\ddot{\chi}+4H\delta u+2\delta\dot{u}}{2M_{g}^{2}}\label{eq:iieq}
%\end{equation}
%\igor{should we have $\delta T_{j}^{i}=\partial_{i}\partial_{j}\chi$
%in all the formulas? i.e. should we consider the most general case $\chi\neq 0$?}
Subtracting the 00 equation from the $ii$ equation, and defining
\be
\label{Xidef}
\Xi\equiv\Phi_{-}+\Psi_{-}
\ee
 gives the following second-order equation for 
the massive fluctuation $\Xi$:
\begin{equation}
\ddot{\Xi} -H_{\rm dS}\dot{\Xi} + \frac{\nabla^{2}\Xi}{a^{2}}+
\left(M^{2}-2 H_{\rm dS}^2\right)\Xi  = J
\label{eq:Xiinhomogeneous}
\end{equation}
with source
\begin{equation}
J={1 \over M_g^2} \left(
-\delta p-\delta \rho+2H_{\rm dS}\delta u-2\delta\dot{u}+\frac{1}{3}\nabla^{2}\chi-H_{\rm dS}a^{2}\dot{\chi}-a^{2}\ddot{\chi} \right) \; ,
\end{equation}
where $\delta p=(1/3)\delta T^k_k$ and
$\delta \rho=-\delta T^0_0$. 
Note that generically, the massive field $\Xi$ is excited
by matter sources because the definition of $\Xi$ in \Ref{Xidef}
contains $g$ fluctuations that do couple to matter. Because of this mixing we cannot think
of $\Xi$ as purely ``new physics'', and in fact it contains a piece
that is present also in GR, as we will see more explicitly below.

We note that if we define a new scalar field
\be
\Pi = {\Xi \over a^2}
\ee
its equation of motion is a (sourced) Klein-Gordon equation
\begin{equation}
\left(\Box-M^{2}\right)\Pi=-J  \; . 
\label{eq:Pieq}
\end{equation}
where the covariant box operator is
\be
\Box  =  -{\partial \over \partial t^2} - 3 H_{\rm dS}{\partial \over \partial t} + {\nabla^2  \over a^2} \; . 
\ee
The identification of the scalar wave equation in massive gravity scenario 
was previously considered in
\cite{Alberte:2011ah}, following older work like \cite{Deser:2001wx}. 

\begin{figure}[h]
\begin{center}
\includegraphics[width=0.8\textwidth]{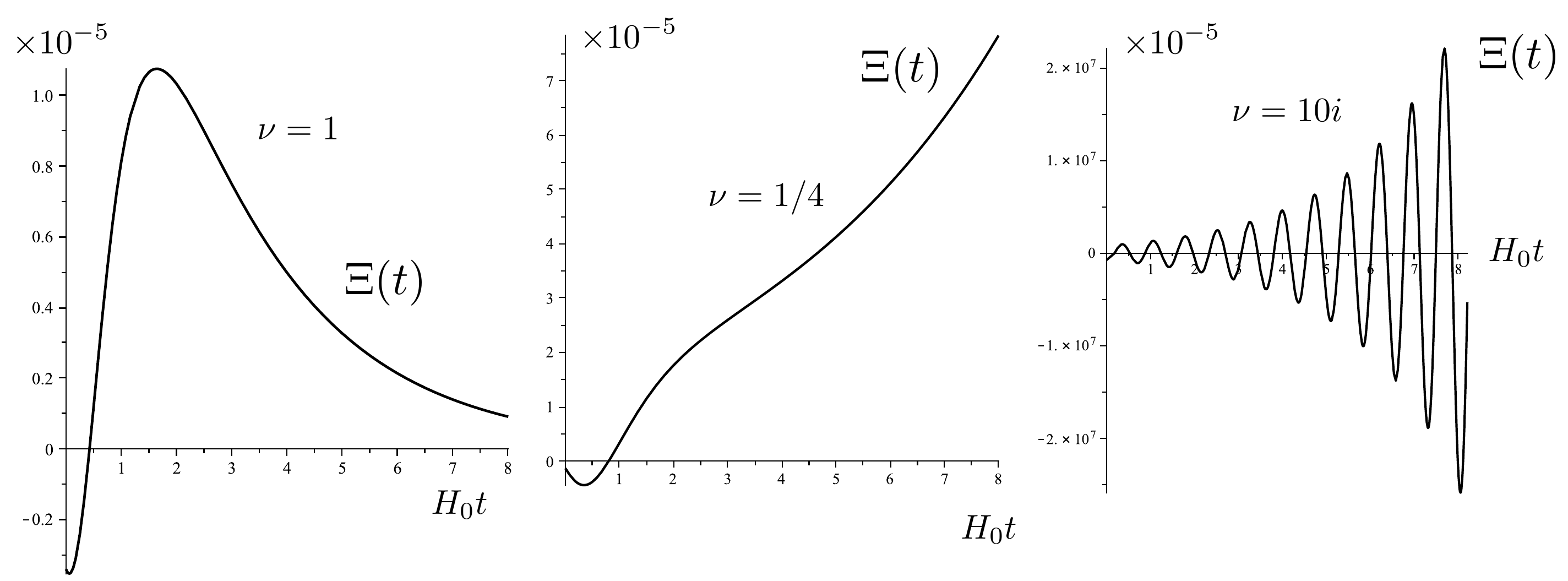}
\caption{The homogeneous massive perturbation
$\Xi(t)=c_J J_{\nu}(x)/\sqrt{x}$, with $x$ from \Ref{xdef}
and $c_J$ a normalization constant, for various $\nu$.
We note that for $\nu <1/2$ or $\nu$ imaginary, this homogeneous mode diverges for late time. However,
for real but small $\nu$ (the second plot) it stays within reasonable values for a several times the age of the universe.
We will only consider $\nu>1/2$ (as in the first plot) in this paper.}
\label{fig:bessel}
\end{center}
\end{figure}

For solving the massive scalar wave equation,
it will be convenient to introduce  a new ``time'' variable $x$ that goes to zero as $t\rightarrow \infty$
\be  \label{xdef}
x ={k \over a_{\rm dS} H_{\rm dS}}= {k \over c_2 H_{\rm dS} } e^{-H_{\rm dS}t }
\ee
with $c_2$ from \Ref{dSsol},
and we introduce a rescaled field
\be
y(x) = \sqrt{x} \, \Xi(x)  \;  , 
\ee
 i.e. $y(t)=\sqrt{k/(c_2H_{\rm dS})} e^{-H_{\rm dS}t/2}\Xi(t)$, 
then $y(x)$ precisely satisfies the inhomogeneous Bessel equation:
\be  \label{Pieq}
y'' + {1 \over x}y' + \left(1-{\nu^2 \over x^2}\right)y = \J(x)
\ee
where the $\nu$ parameter (the order of the Bessel function) is
\be  \label{nueq}
\nu^2 = {9 \over 4} - {M^2 \over H_{\rm dS}^2} \; . 
\ee 
Here $M$ is given in \Ref{eq:effectiveparameters}
and the source $\J(x)$ is found from the massless sector,
i.e.\ from \Ref{rhodS} and \Ref{udS}. Setting $\chi=0$ and $\delta p=0$
it simplifies to
\be  \label{source}
\J(x) = {2 c_2 C_1  H_{\rm dS}  x^{3/2} \over k}
+ {20 c_2^3 C_2  H_{\rm dS}^3 x^{3/2} \over k^3} +{2 c_2^3 C_2  H_{\rm dS}^3  x^{7/2} \over k^3} \; . 
\ee
We now proceed to write down the solutions of \Ref{Pieq} for $y(x)$ and use them to recover 
$\Pi$ of \Ref{eq:Pieq}. We
will only consider $M^2<(9/4)H_{\rm dS}^2$ here,
such that $\nu$ in \Ref{nueq} is real; the solution for $\nu$ imaginary
is discussed somewhat further in the appendix. 
We give representative plots of the corresponding $J_{\nu}$ Bessel functions
in fig.\ \ref{fig:bessel}. 

\subsection{Solutions of massive wave equations in dS: real case}

\begin{figure}[h]
\begin{center}
\includegraphics[width=0.8\textwidth]{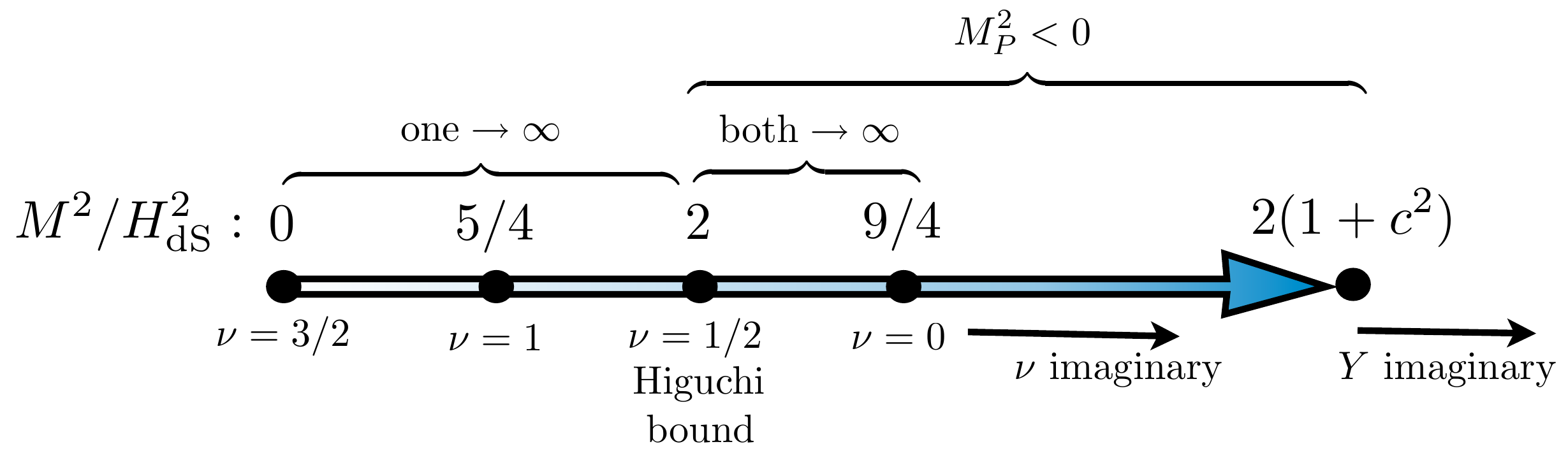}
\caption{Summary of some interesting values of the mass parameter $M$
and the corresponding $\nu$ values from \Ref{nueq}.
We indicate when only one homogeneous mode goes
to infinity at late time, or when both modes do, and
when $\nu$ becomes imaginary (which is not 
a problem in itself, but see fig \ref{fig:bessel})
and when $Y$ and $M_P^2$ become negative
(which are big problems for the background in the current model).}
\label{fig:M}
\end{center}
\end{figure}

As we argue in the appendix, the solution of \Ref{Pieq}  is
\be  \label{ysol}
y(x)= c_{{\rm J}} J_{\nu}(x) + c_{{\rm Y}} Y_{\nu}(x) + 
y_{\rm p}(x)
\ee
where the particular solution $y_{\rm p}$ is obtained from \Ref{source} and
\Ref{genericpart} as
\be
y_{\rm p}(x)
 = {2 c_5 C_1  H_{\rm dS} s_{5/2,\nu}(x) \over k}
+ {20 c_5^3 C_2  H_{\rm dS}^3 s_{5/2,\nu}(x) \over k^3} +{2 c_5^3 C_2  H_{\rm dS}^3  s_{9/2,\nu}(x) \over k^3} \; . 
\ee
The $s_{\mu,\nu}(x)$ are 
special functions that are solutions to the inhomogeneous  Bessel equation with power-like source,
also called Lommel functions, and are
 defined
in appendix \ref{app:bessel}. By the asymptotics given in that appendix,
this $y_{\rm p}$ vanishes as $x^{7/2}$ for $x\rightarrow0$ (late times). 
If we now fix $c_{\rm Y}=0$ in \Ref{ysol}, the homogeneous solution $J_{\nu}(x)$
is proportional to $x^{\nu}$
for $x\rightarrow0$. The parameter $\nu$ depends on the mass parameter $M$ and is given in \Ref{nueq}.
We see that since $\nu<7/2$ (compare fig.\  \ref{fig:M}), 
the homogeneous solution will be leading (i.e. its leading power of $x$ will be {\it lower}) compared to the particular solution, at late times. Of course, we can also turn off the homogeneous solution at will 
by setting $c_J=0$, but the particular solution always remains
in the massive scalar $\Pi$. 

In fig.\ \ref{fig:bessel}, some of the solutions 
of the homogeneous equation blow up at late time. 
Moreover, this is just the $J_{\nu}$ solution; the $Y_{\nu}$ solution
blows up at late times for all $\nu$ in our range.
This is not necessarily
a problem for the theory, though it is a problem for our approximation; 
all it means here is that 
if these modes would be excited, linear perturbation theory breaks down in the
(possibly distant) future.
Even if nonlinear fluctuations around
this background did produce some instability in  in the future, we do not believe this is
relevant to our phenomenological objectives in this paper,
simply because the far future is more of an auxiliary device
here than a region of interest. For example, 
one can picture a scenario where  the current model in Hassan-Rosen bimetric theory
is replaced by another effective theory at late times, where the growing solution is matched
to a decaying solution in the new theory. It would be interesting
to learn that this is not possible and actually
the presence of these modes rule out the theory
for some values of $M$, but at the moment this is not clear to us.
See also the conclusions for more comments on this.

We summarize in figure \ref{fig:M} why $\nu>1/2$ seems
the only reasonable choice in the current model.
%\subsection{Solutions of massive wave equations in dS: complex case}
%Now for the case when $\nu^2<0$ in \Ref{nueq}. The homogeneous solution to this equation is given
%in terms of Hankel functions by 
%\[
%\Xi^{H}\left(x,t\right)=\int d^{3}k\left[C\left(k\right)e^{\frac{Ht}{2}}H_{\nu}^{1}\left(\frac{k}{H}e^{-Ht}\right)e^{ikx}+C\left(k\right)^{*}e^{\frac{Ht}{2}}H_{\nu}^{1}\left(\frac{k}{H}e^{-Ht}\right)^{*}e^{-ikx}\right].
%\]
%
%
%The homogeneous solution will have a relaxation time
%
%\[
%\tau_{\tiny\mbox{rel}}^{H}=\cdots
%\]
%where $H$ stands for 'homogeneous'. The particular solution can not
%be given in an analytic form, but since the source term has a relaxation
%time of
%\[
%\tau_{\tiny\mbox{rel}}^{S}=\cdots
%\]
%we expect the particular solution to have the same relaxation time.
%This is also confirmed numerically. 
%

\subsection{Parameter values}

%The relevant scale, for $M^{2}>H_{\rm dS}^{2}$, is 
%set by which region $k^{2}$ belongs to. There are three possible
%scenarios:
%\begin{enumerate}
%\item The wavelength of the perturbation is larger than the dS Hubble radius,
%and hence also larger than the length scale set by $M^{2}$. 
%\item The wavelength of the perturbation is smaller than the dS  Hubble radius
%but larger than the length scale set by $M^{2}$.
%\item The wavelength of the perturbation is smaller than both the dS  Hubble
%radius and the length scale set by $M^{2}$.
%\end{enumerate}
%Fixing the value of $M^{2}$ relative to $H_{\rm dS}^{2}$, we  then choose
%the value of $k^{2}$ following the three possible scenarios
%given above. 
We now have four integration constants that need to
be specified: First, the constants $C_1$ and $C_2$ of \Ref{Psiplus}
for $\Psi_{+,\rm dS}$,
which we can think of as initial conditions (ICs) for $\Psi_{+,\rm dS}$ and $\dot{\Psi}_{+,\rm dS}$,
i.e. the gravitational potential perturbation of the massless sector. 
Or equivalently, we can think
of them as ICs on the values
(but not the derivatives) of $\delta\rho_{\rm dS}$ and $\delta u_{\rm dS}$.
Second, we have the 
the constants $c_{\rm J}$ and $c_{\rm Y}$  from
the homogeneous solution for $\Xi_{\rm dS}$,
which we can think of as ICs on  $\Xi_{\rm dS}$ and
$\dot{\Xi}_{\rm dS}$, i.e. the {}``wave-like'' field in the massive sector.
Thus in general we have four parameters $\{ C_1, C_2, c_{\rm J}, c_{\rm Y}\}$.

For convenience we give the mass $M^{2}$ and the comoving
wavenumber $k^{2}$ in terms of the  Hubble
parameter value $H_{\rm dS}=\sqrt{\Omega_{\Lambda}}H_0$ that is asymptotically approached in the future, i.e. not in terms of 
the Hubble parameter value $H_0$ of today. 

%
%\subsection{de Sitter Higuchi bound}
%
%Instabilities?

\subsection{Conclusions in dS: nothing new
for cosmological geometrical probes}
Because the perturbation equations for the ``$+$" subscript fields, 
eqs.\ \Ref{massless1}-\Ref{massless4},
are identical to the GR equations, and the 
``$-$" subscript fields only appear in the
massive equations  \Ref{massive1}-\Ref{massive4},
and never appear in the GR-like equations at this order, the matter perturbations 
and hence the linear growth of structure
will be identical to that seen in general relativity. It is important for this conclusion  that the matter perturbations are completely determined by the $00$ and $0i$ components of Einstein's equation, which are of course constraints. In general in massive gravity this may not be the case; see the conclusions
for related comments. 

We also note that the argument in the previous paragraph leaves much to be desired from a phenomenological
standpoint. For example, as light rays propagate along geodesics of the $g$ metric, 
presumably one cannot form linear combinations of fluctuations
in the electromagnetic field as we have done here
that look identical to the GR equations, even in pure de Sitter, so one could expect
that bending of light and similar local experiments will be grossly affected even
where the cosmological matter perturbations are not. However, 
we emphasize that local solutions are not yet well understood,
and we follow the practice of \cite{vonStrauss:2011mq} of lumping
these unknowns into $M_P$,
that is never directly relevant in global equations. 

\section{Quasi-de-Sitter space}
\label{qdS}
We have seen that bimetric cosmological gravitational perturbations in de Sitter space 
are identical to those of GR. For this and other reasons, it is of interest to consider a universe that represents
a small devation from
from  the de Sitter {\it background}, and then consider cosmological perturbation 
theory in this  slightly generalized background. This kind of approach works well in inflation, but
it is not often used for late-time cosmology where the
presence of matter complicates things. We will see some of these complications, and why 
the quasi-de-Sitter approach is still
useful for our purposes. 
For some background on quasi-de-Sitter, it is useful to consult 
a review of inflation, e.g.\ \cite{Baumann:2009ds}.

\subsection{Background}
We define quasi-de-Sitter space as near-exponential expansion of the scale factor\be   \label{slowroll}
\tilde{\epsilon} \equiv -{\dot{H} \over H^2}  > 0
\ee
where as usual $H(t)$ is defined as $H=\dot{a}/a$,
and the square in the denominator makes $\tilde{\epsilon}$ dimensionless. For pure de Sitter space,
 $H$ is strictly constant so $\tilde{\epsilon} =0$ in pure dS.
 (The reason for the tilde will become apparent shortly.)
We remind the reader that 
also in inflation
$\epsilon$ is often first defined in terms of the geometry, just like in \Ref{slowroll}. The inflationary slow-roll parameters in inflation are then given in terms of some scalar field potential, for which we have no direct analog here. 

It is a simple matter to compute $-\dot{H}/H^2$ from the Friedmann equation \Ref{Fried3} to obtain
\be \label{epsphys}
\tilde{\epsilon} \; \approx \; {3 \over 2}{\Omega_m \over \Omega_{\Lambda} a^3} 
\ee
where we have used the zeroth-order relation $H(t) \approx H_{\rm dS}$.
We see that $\tilde{\epsilon}$ encodes the fraction of matter in a universe
dominated by dark energy. Clearly, this should be a small parameter in the future, and
as we approach matter domination, the quasi de Sitter approximation breaks down
in the past. \footnote{In fact,
the expansion in e.g.\ Ch.\ 8  of Weinberg \cite{Weinberg:2008zzc} using the correction factor $C(x)$
is the inverse expansion of this,
there it is the ratio of dark energy to matter, which breaks down around present
and more severely in the future.}

We split the Robertson-Walker scale factors of the two metrics into products
of de Sitter scale factors and correction factors:
\be
a(t) &=& q_a(t) a_{\rm dS}(t) \; ,  \qquad q_a(t\rightarrow \infty) = 1 \\
Y(t) &=& q_Y(t)Y_{\rm dS}(t)  \; ,  \qquad q_Y(t\rightarrow \infty) = 1
\ee
so that the functions $q_a(t)$ and $q_Y(t)$ capture the ``quasi-ness'' of the expansion.
%We 
%can determine the expansion coefficients $\epsilon_{g,i}$
%from the $H^2$ equation, the expansion coefficients $\epsilon_{m,i}$
%of $\Omega_{m}$ from the continuity equation, 
%then we determine the expansion coefficients $\epsilon_{f,i}$ from the equation for $\Upsilon^2$. 
Let us consider the exact solution \Ref{exact}, \Ref{dSsol} for the two-component fluid, then we 
can extract the quasi-ness for $a(t)$ as
\be  \label{qex}
q_a(t) = {a(t) \over a_{\rm dS}(t)} = 2^{2/3} {\sinh\left({3 \over 2}H_{\rm dS} t\right)^{2/3} \over \exp{H_{\rm dS} t}}
=\left(1-{1 \over 6}\epsilon(t)\right)^{2/3}
\ee
exactly, where now we {\it define} 
\be  \label{eexpl}
\epsilon(t)\equiv 6e^{-3H_{\rm dS} t}
\ee
which agrees with \Ref{epsphys} to lowest order (hence the tilde in \Ref{epsphys}).
For some numbers
to keep in mind,
$\epsilon \sim 0.01$
at $H_0t \sim 2.5$,
$\epsilon \sim 0.1$
at $H_0t \sim 1.6$
and
$\epsilon \sim 0.5$
at $H_0t \sim 1$.
We will prefer to stay at $\epsilon<1$,
which limits us to $H_0t > 0.7$ as a matter of principle.
(In actual examples, we will find greater limitations than
this.)
For small $\epsilon$, we can expand the quasi-ness of $a(t)$ in $\epsilon$:
\be  \label{qexp}
q_a(t) = 1-{1 \over 9} \epsilon(t) -{1 \over 324}\epsilon(t)^2 +  \ldots
\ee
%It may be useful to contrast this with 
%the alternative approach of writing a series
%expansion in $H_0(t-t')$ starting at some fiducial time $t'$.
%This amounts to expanding the exponential factor
%in \Ref{qexp}, so it breaks down completely when $3c_2(t-t') \sim 1$,
%which is about $H_0(t-t') \sim 1/(3c_2) \sim 0.4$,
%i.e.\ 
%about 5 Gyr of evolution,
%so for a reasonable approximation 
%we could hope to go one or two gigayears
%back from a given fiducial point $t=t'$,
%and this will typically not let us evolve back to the present
%with good accuracy. 
%In practice these estimates turn out to be slightly
%too conservative, but the main point remains
%that the linear approximation in $H_0(t-t')$
%is not good around the present and earlier.
%This is illustrated in figure \ref{orders}.
%We will have some use for both kinds of approximation. \mar{Maybe.}
We can then easily compute the (square of the) Hubble function:
\be
H^2 \equiv {\dot{a}^2 \over a^2} = H_{\rm dS}^2\left(1+{2 \over 3}\epsilon
+{2 \over 9}\epsilon^2 \right) 
\ee
leading to
\be
 -{\dot{H} \over H^2} = \epsilon(t) + \ldots  \; . 
\ee
Note that because we defined
$\epsilon(t)$ as \Ref{eexpl}, this is not {\it exactly} $\tilde{\epsilon}(t)$
of \Ref{slowroll}. This distinction is
one of convenience and merely amounts to a rearrangement
of higher-order perturbation theory in the ``true'' quasi-de Sitter
parameter  $\tilde{\epsilon}$. 

For the $f$ metric
one can argue similarly. 
From \Ref{Y2} 
and with the dS solution in \Ref{dSsol}, 
\be  \label{qYexp}
Y(t) &=& Y_{\rm dS}
\left(1-{c_Y \over 9}\epsilon-{c_{Y,2} \over 324}\epsilon^2 + \ldots\right)
\ee
where 
\be
c_Y &=& {M^2 + 4(1+c^2)H_{\rm dS}^2 \over M^2-2(1+c^2)H_{\rm dS}^2}  \label{cY} \\
c_{Y,2} &=& {M^4 + 44(1+c^2)M^2 H_{\rm dS}^2-20(1+c^2)^2H_{\rm dS}^4
\over (M^2-2(1+c^2)H_{\rm dS}^2)^2} \; . 
\ee
This qdS expansion of the scale factor $Y$ captures the  loss of proportionality 
between $Y(t)$ and $a(t)$ as we 
leave the pure dS regime and enter the quasi-de Sitter regime.
(Note that the constant $c_Y$ is never unity.)
We summarize the results for qdS expansion coefficients for the
various 
derived background quantities
in appendix \ref{app:repl}
at linear order, which is all we will use explicitly in this paper.

\begin{figure}
\begin{center}
\includegraphics[width=0.45\textwidth]{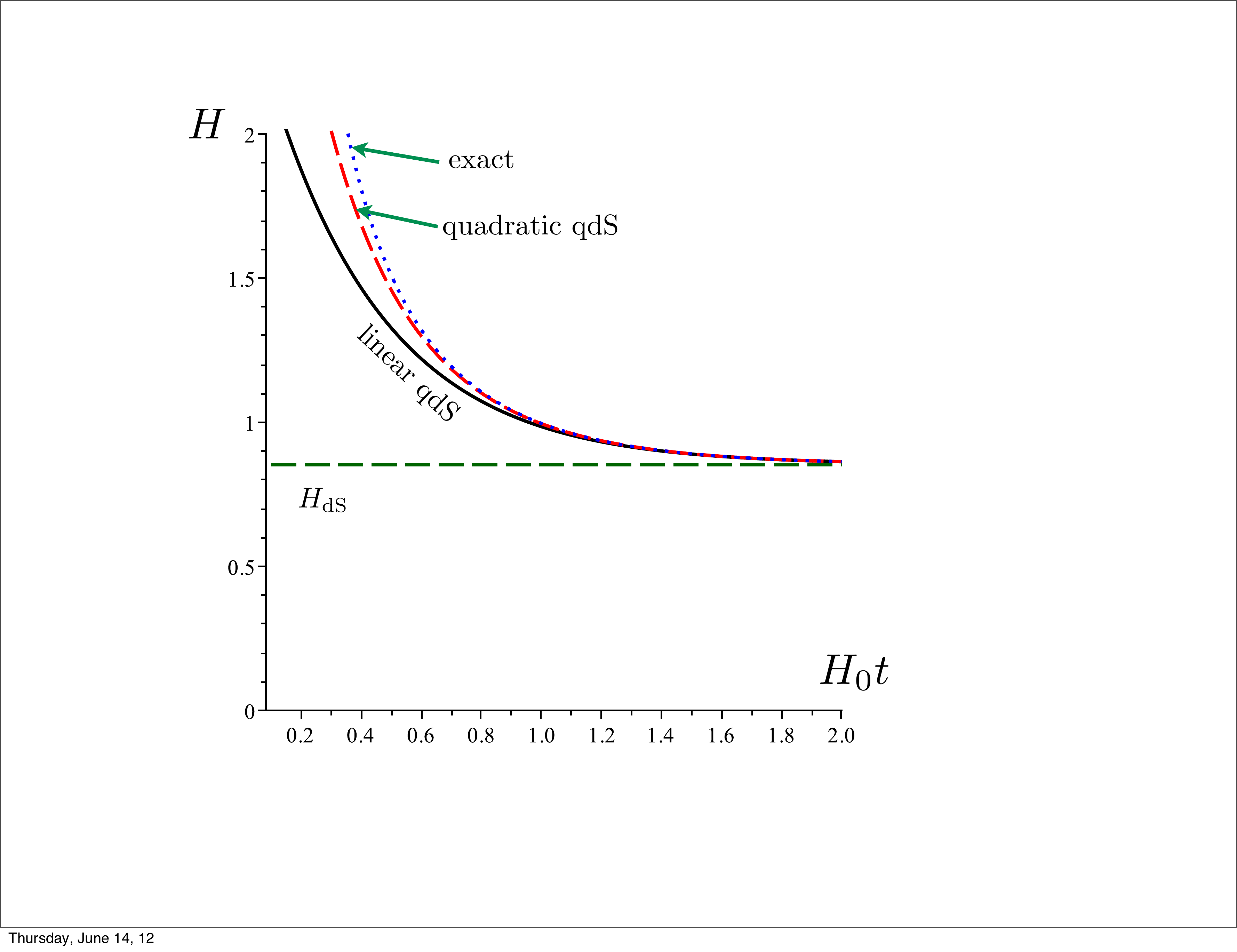} \hspace{2mm}
\includegraphics[width=0.45\textwidth]{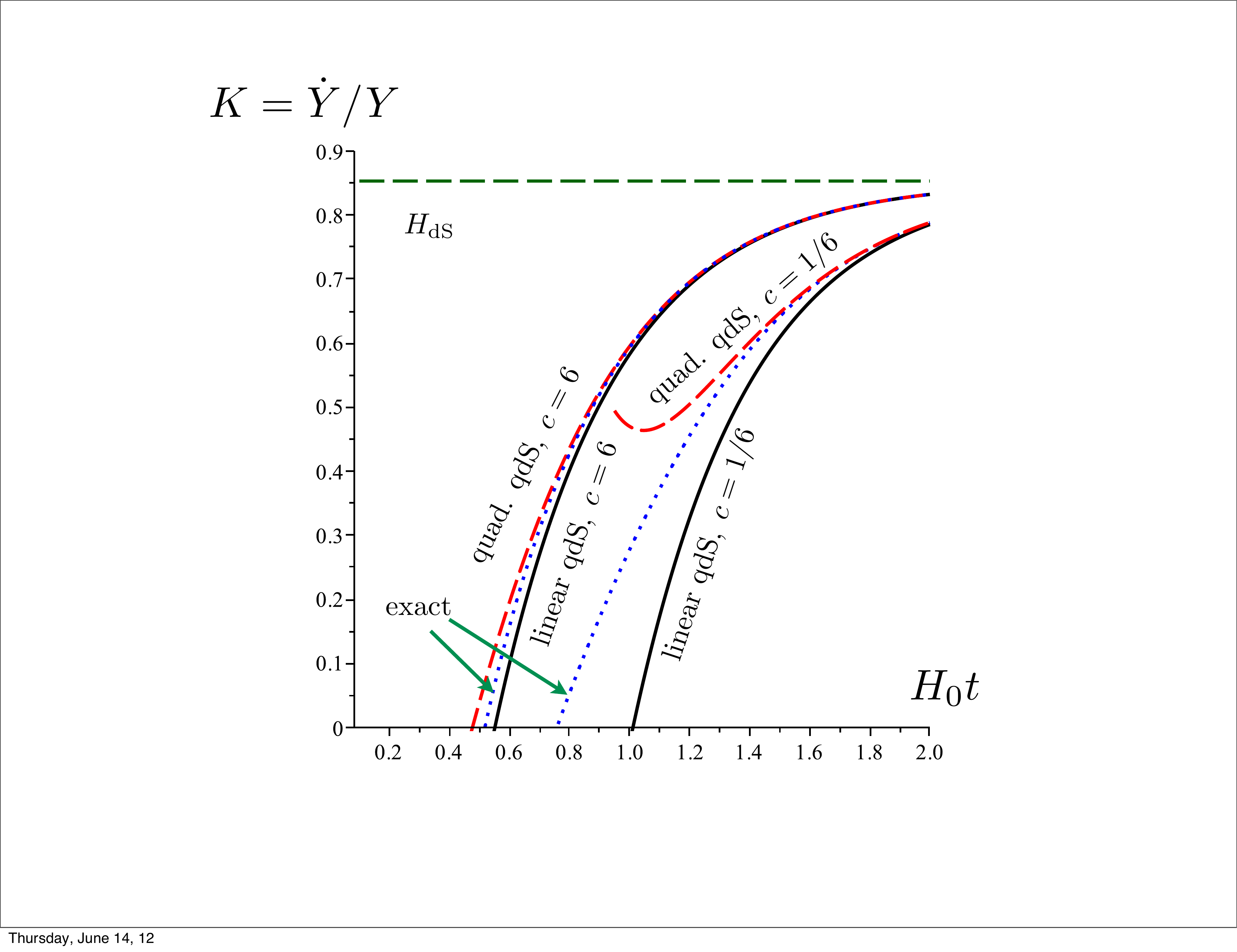} \\
\caption{Background solutions for the Hubble
functions $H=\dot{a}/a$ and $K=\dot{Y}/Y$, with the latter for 
 $c=6$ and $c=1/6$, cf.\ \Ref{constprop}.
 (Here the $c=1/6$ curves are included for illustration only, we never use 
 values for $c$ this low).
Linear (in $\epsilon$) qdS in black,
quadratic  qdS in dashed red,
 exact solution in dotted blue,
and $H_0t \sim 1$ is roughly present. We see that
for times $H_0 t \gtrsim 0.7$, qdS remains a good approximation
for the
$g$ background, and also for the $f$  background for $c=6$.}
\label{fig:H}
\end{center}
\end{figure}

%Or, if we give up some accuracy, we can expand $\epsilon$ to
%obtain constant coefficients:
%\be
%H &=& c_2 \left(1+ {1 \over 3}\epsilon(t) \right)
%=c_2\left(1+ {1 \over 3}(\epsilon(t')+\dot{\epsilon}(t')(t-t')+\ldots) \right)\\
%&=&H_{\rm dS} +  H_{\epsilon}(t-t') + \ldots
%\ee
%where $H_{\rm dS} = c_2\left(1+ {1 \over 3}(\epsilon(t')\right)$
%and
%$H_{\epsilon}=(c_2/3)\epsilon'(t')$
%are both constants. 
%Since $\dot{\epsilon}(t)=-3c_2\epsilon(t)$, we see that $\dot{H}=
%H_{\epsilon}=-\epsilon(t')H_{\rm dS}^2$
%at this order, as expected.

\subsection{Perturbations in qdS}
We write the general perturbation equations
in section \ref{eom} as $D_{mn}  \, \phi^n=\J_m$  for a collection of fields $\phi^m$
enumerated by $m=1,\ldots,8$ and a differential operator $D_{mn}$
and sources $\J_m$.
We organize the expansion as follows:
\be
D_{mn} = D^0_{mn} + \epsilon D^{\epsilon}_{mn} \; , \quad
\phi^m =  \phi^m_0 + \epsilon \phi^m_{\epsilon} \; , \quad
\J_m = \J_m^0 + \epsilon \J_m^{\epsilon} \; , 
\ee
from which we write the first order equation as 
\be 
D^{0}_{mn} (\epsilon \phi^n_{\epsilon}) &=& \epsilon \J_m^{\epsilon} -\epsilon \,  D^{\epsilon}_{mn} \, \phi^n_{0} \\
&=&\epsilon (\J_m^{\epsilon} +\tilde{\J}_m^{\epsilon})
\label{Dlin}
\ee
using the zeroth order de Sitter equations  $D^0_{mn}  \phi^n_0=\mathcal{J}^0_m$,
and neglecting quadratic order in $\epsilon$. 
As expected, the zeroth order (pure de Sitter) fields
 act as additional sources $\tilde{\J}_m^{\epsilon}$
for the first order quasi-de Sitter fields. 
It is useful that from this vantage point, the differential 
operator on the left-hand side of \Ref{Dlin} is the unperturbed de Sitter differential operator $D^{0}_{mn}$. 

Note that  \Ref{Dlin} contains terms
with the time derivative $\dot{\epsilon}$
of our perturbation parameter. In principle
by using \Ref{eexpl} we can
express these entirely in terms of $\epsilon$,
and the latter would then drop out
of the equation. In practice, it
is convenient to work with the unperturbed dS differential operator
also in qdS,
so we will not substitute in for $\dot{\epsilon}$
and instead simply write
\be
\phi^m =  \phi^m_{\rm dS} +  \phi^m_{\rm qdS} \; \; . 
\ee
The price to pay for this convenience is that viewed as expansions
in $x\sim e^{-H_{\rm dS}t}$, it is not guaranteed
that every term in $\phi^m_{\rm qdS}$ is suppressed
compared to every term in $\phi^m_{\rm dS}$,
and in fact it will generically not be the case,
but  the
overall   series in  $x$ does display relative suppression.
We summarize this fact in table \ref{Deltatab}. 
For more explicit comments on this issue in the simpler setting of
ordinary GR, we refer the reader to appendix \ref{app:GR}. 

It would be useful to know when the linear (in $\epsilon$) qdS approximation is valid to
some some prescribed accuracy for the perturbations. However, we currently do not have 
perturbations in an exact solution
to compare to in bimetric theory, so it is hard to be absolutely precise about
this (and if we did have an exact solution, the question would be rather pointless).
As a first check, we compute the relative errors
of the linear and quadratic qdS approximations
versus the exact solution in
cosmological perturbation theory in pure Einstein gravity
in the aforementioned appendix \ref{app:GR}.
As a second check, we have performed some preliminary  analyses
of the {\it quadratic} (in $\epsilon$) qdS approximation also in bimetric theory,
in particular how  the
bimetric perturbations in the
 quadratic qdS approximation differ from the linear qdS approximation (typically if second order
perturbation theory produces significant changes, perturbation theory has broken down).
To be clear, for the purposes of this paper we only 
use the quadratic qdS approximation for auxiliary checks and we do not display it in plots.
We find numerically that the linear approximation in $\epsilon$
is good to about $10$ to $30\%$ for the bimetric perturbations (depending on the field)
for  $H_0t \gtrsim 1.3-1.5$. This will be indicated by shading
the region below this in the plots. 

From the good accuracy of the qdS background in section \ref{fig:H} one
could have hoped that the linear qdS perturbations would
have extended further back than $H_0t \gtrsim 1.3-1.5$, since the
future is of no direct use for phenomenology,
but there was of course no guarantee that this would be the case.
On the good side, since going to quadratic order seems to give some improvement in our preliminary analyses, we believe that the qdS approximation
at higher orders will
be useful also for phenomenological purposes,
and  not only as indirect checks of numerical solutions of the perturbation equations
in the exact background.

\subsection{Parameters}
We use wavenumbers $k=10H_{\rm dS}$, $k=(5/2) H_{\rm dS}$
and $k=(1/2) H_{\rm dS}$ as representative cases,
the latter only for internal checking of the analytics,
as we will describe later.
For $k=(5/2) H_{\rm dS}$
we will impose 
the following values on the matter
perturbations at $H_0t' =1.5$. (One would have liked to do this at present $t=t_0$,
but the qdS approximation needs to be valid in the region
where we set initial conditions.)
We  obtain the values from GR (see appendix \ref{app:GR}):
\be \label{mattervalues}
{\rho(t') \over M_P^2} = 8.93 \cdot 10^{-5} 
\; , \quad  
{u(t') \over M_P^2}= 0.72 \cdot 10^{-5} 
\ee

For $k=(1/2) H_{\rm dS}$, we have instead
\be \label{mattervalues2}
{\rho(t') \over M_P^2}= 2.43 \cdot 10^{-5} 
\; , \quad  
{u(t')\over M_P^2} = 0.72 \cdot 10^{-5}  \; . 
\ee
We view these as having roughly 10\% accuracy.
It is nontrivial to extract the values
directly from data, but doing so would be useful
in a more phenomenological analysis,
rather than comparing directly to Einstein gravity, since we are of course modifying gravity.

\subsection{Analytical series solution}
The qdS equations are more complicated than the dS equations,
but as the differential operator  on the left-hand side of \Ref{Dlin} is the unperturbed de Sitter differential operator,
the general strategy for solving the differential equations is the same. 
In particular we again find a massive inhomogenous Bessel equation
for $\Xi_{\rm qdS}$, just with more complicated sources. 
We will not turn on this homogeneous solution,
since if we did, it should have been included at dS order (if this is unclear,
it may help to consult our analogous comments in GR in appendix \ref{app:GR}).
We introduce $y=x^{1/2}$, cf.\ \Ref{xdef}.
We compute series expansions of the right hand sides
of all the perturbation equations to see 
which powers of $y$ actually occur, and arrive at the following
series ansatz:
\be
\Phi_{+} &=& \Phi_{(7)}\,  y^7 + \Phi_{(8)}\,  y^8   + \Phi_{(11)}\,  y^{11}  + \Phi_{(12)}\,  y^{12}  \\
\Psi_{+} &=& \Psi_{(7)}\,  y^7 + \Psi_{(8)}\,  y^8 + \Psi_{(11)}\,  y^{11}  + \Psi_{(12)}\,  y^{12}  \\
\delta \rho&=& \delta\rho_{(7)}\,  y^7 + \delta\rho_{(8)}\,  y^8  + \delta\rho_{(11)}\,  y^{11}  + \delta\rho_{(12)}\,  y^{12}  \\
 \delta u &=& \delta u_{(7)}\,  y^7 + \delta u_{(8)}\,  y^8  + \delta u_{(11)}\,  y^{11}  + \delta u_{(12)}\,  y^{12}  \\
\Xi &=& \Xi_{(7)}\,  y^7 + \Xi_{(8)}\,  y^8   + \Xi_{(11)}\,  y^{11}  + \Xi_{(12)}\,  y^{12}  \\
\Phi_{-} &=& \Phi_{-(7)}\,  y^7 + \Phi_{-(8)}\,  y^8 + \Phi_{-(11)}\,  y^{11}  + \Phi_{-(12)}\,  y^{12}  \; . 
\ee
(The ${\mathcal F}$ and ${\mathcal B}$ fields are determined from these, as before.)
The explicit expressions for the coefficients obtained in this way
are not terribly illuminating so we do not present them in full, 
but to give an idea of what they look like  for our fixed parameter values
(in  particular $\nu=1$),
we find coefficients of the rather manageable form
\be
\Phi_{(7)} &=&= {6c_2 H_{\rm dS}^3 \over 5k^3 }\cdot 
{48 M^4 -176 M^2 H_{\rm dS}^2 + 205 H_{\rm dS}^4 
\over (M^2-74 H_{\rm dS}^2)(M_{\rm dS}^2-2 H_{\rm dS}^2)}\cdot c_J
= {128 \over 97} {c_2^3 H_{\rm dS}^3 \over  k^3} \cdot c_J \\
\Phi_{(8)} &=& -{8 c_2^4 H_{\rm dS}^4 \over 3 k^4 } 
{M^2 + 34 H_{\rm dS}^2  \over  (M^2-74 H_{\rm dS}^2)(M_{\rm dS}^2-2 H_{\rm dS}^2)}
\cdot C_1
= -{376 \over 291} {c_2^4 H_{\rm dS}^4 \over  k^4} \cdot C_1
\ee
and so on.
For our  parameters,
$2(1+c^2)=74$, so both denominators
are $(M^2-2(1+c^2)H_{\rm dS}^2)(M^2-2H_{\rm dS})$,
i.e. the expansion breaks down not only for early times
but also for two of the distinguished mass parameter values in figure \ref{fig:M}. 
The breakdown point 
$M^2=2(1+c^2)H_{\rm dS}^2$ could have been anticipated from 
the background expansion \Ref{cY}. 

For $k=(1/2) H_{\rm dS}$, we find 
that the present time $t=t_0$ corresponds to $x= 0.47$ (see \Ref{xdef}), and  for $k=(5/2) H_{\rm dS}$
we find $x=2.37$. So for the larger $k$
values, the more   phenomenologically interesting ones,
we expect this analytic version of the qdS expansion to break down,
even when a numerical qdS approach would still be valid.
(However, the analytics may be somewhat better than this,
since many terms are actually expansions in $x/k$, which is independent of $k$.)
Therefore we focus on $k=(1/2)H_{\rm dS}$ 
when we use the analytical series solution.
\begin{table}
\begin{center}
\begin{tabular}{|c|c|c||c|c|c|}\hline
field & $\Delta_{\rm dS}$ & $\Delta_{\rm qdS}$ &
field & $\Delta_{\rm dS}$ & $\Delta_{\rm qdS}$   \\  \hline \hline
$\Psi_+$ &   1  &  7/2 & $\Xi$ & 3 & 4 \\ \hline  
$\Phi_+$ &   1  &  7/2 & $\Phi_-$ & 3 & 4 \\ \hline  
$\delta\rho$ &   3  &  7/2 & ${\mathcal B}$ & 5 & 5  \\ \hline  
$\delta u$ &   3  &  7/2  & ${\mathcal F}$ & 4 & 4 \\ \hline  
\end{tabular}
\caption{Leading expansion powers in the analytical solution.
The notation means that the leading term at late time is $x^{\Delta}$.}
\label{Deltatab}
\end{center}
\end{table}
In all, the lessons we learn from the analytical 
solution is that there are certain degenerate special parameters,
and we can quantify when a given approximation 
breaks down at least for small wavenumber $k$. None of this will be evident in the following,
since we have already identified useful parameter values
and we will not bother to show plots comparing the analytical and numerical results, we will just state here that they agree to the extent we expect them to.
Perhaps the most important use is as
cross-check with the numerics for low $k$. 
We now turn to the numerics.

\subsection{Numerical solution: general}

Naively we would expect
that the energy density would receive a slight positive correction since
going away from  pure de Sitter expansion means that there is less
expansion and the friction term due to the expansion is thus smaller.
But in bimetric gravity the situation is more involved since both
the massless and massive background sector will contribute to the
correction. For example, the initial conditions set for the massive wave (i.e.
the two integration constants for the homogeneous solution) will 
affect the correction to the energy density.

One way to fix parameters would be to use $\rho$, $u$ and $\dot{\rho}$ 
 at present
to fix ICs $C_1$, $C_2$ and $c_{\rm J}$ (we always set $c_Y=0$).
Another way is to fix $c_{\rm J}=0$
and fix $C_1$, $C_2$ from
$\rho$, $u$ at present, which is what we will do here (with the exception
of figure \ref{deltacj}).
 For the values \Ref{mattervalues2} for $\delta \rho$ and $\delta u$,
 and $k=(5/2) H_{\rm dS}$, we obtain
\be
C_1=-2.19 \cdot 10^{-5}, \quad C_2=3.33 \cdot 10^{-5} \; , \quad
c_{\rm J} = 0  \; . 
\ee
where we have normalized the gravitational potential
as in GR (see appendix \ref{app:GR}). 

\subsection{Numerical solution: gravitational potential}
\begin{figure}
\begin{center}
\hspace{-7mm}\includegraphics[width=0.45\textwidth]{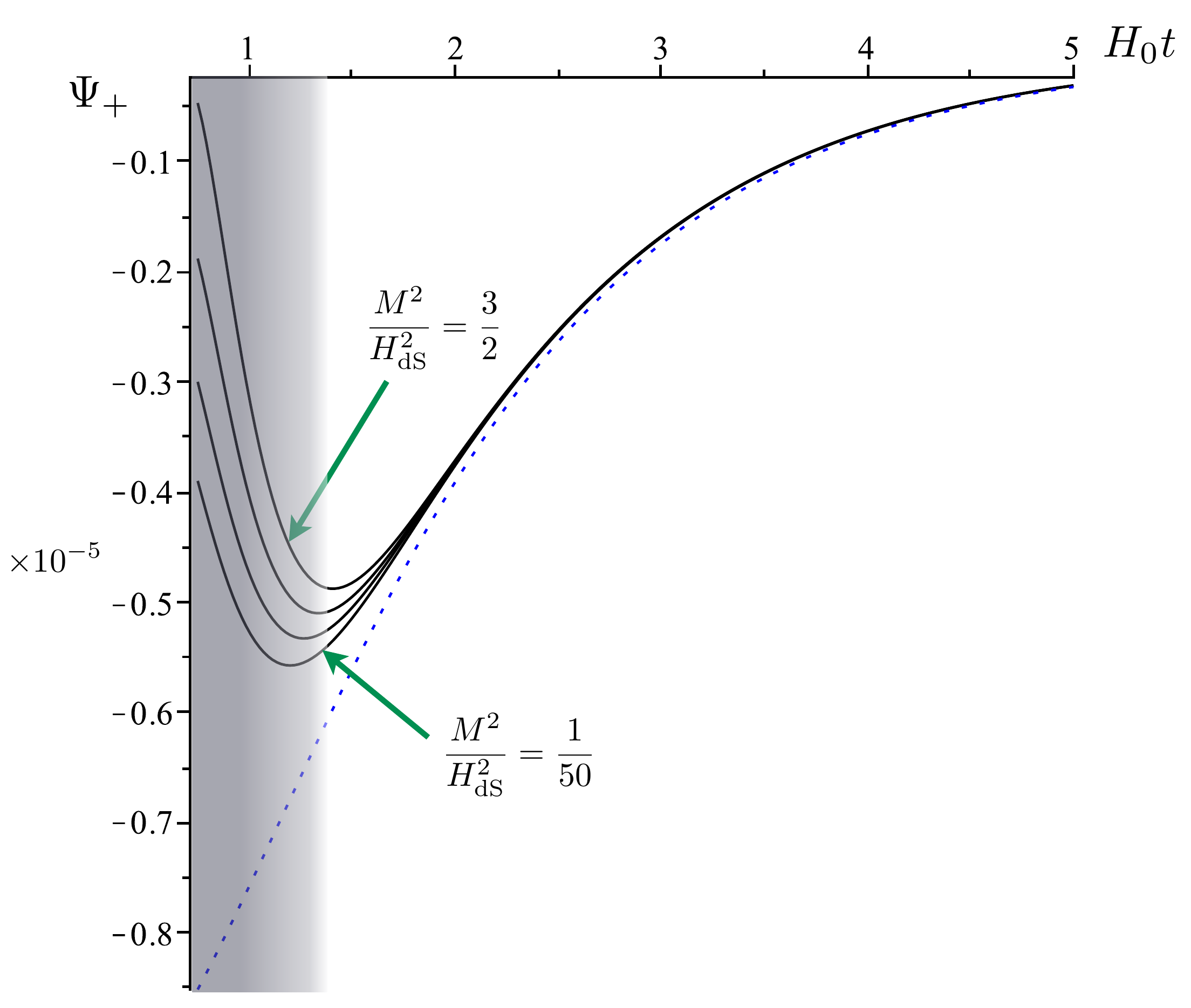}\hspace{2mm}
\includegraphics[width=0.45\textwidth]{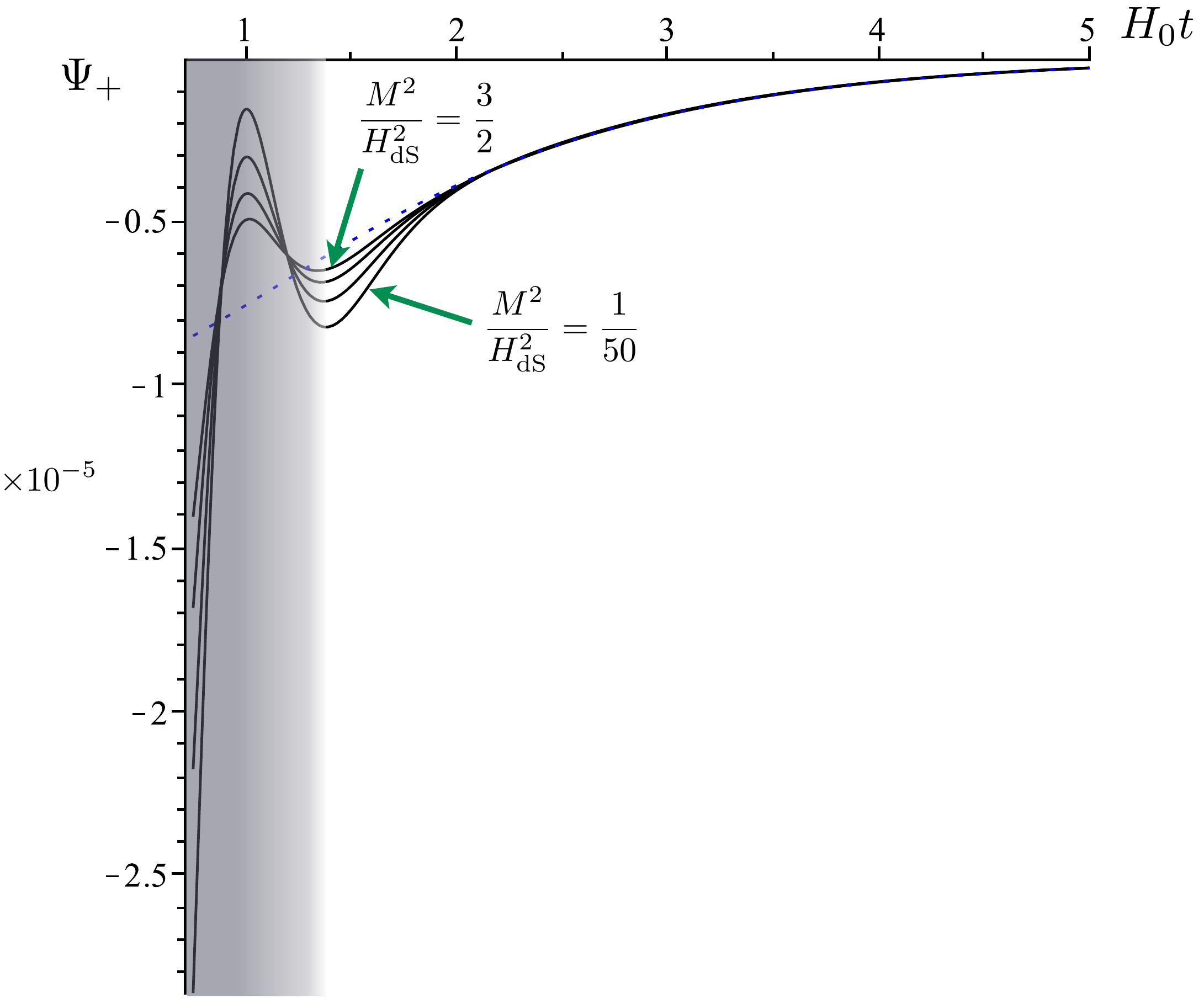}
\caption{Numerical qdS plots for the gravitational potential 
$\Psi_+$ for $k=(5/2) H_{\rm dS}$ (left panel) 
and $k=10 H_{\rm dS}$ (right panel). 
and $M^2/H_{\rm dS}^2 =  \{ 1/50, 4/5, 5/4, 3/2 \}$.
The shaded area is our estimate for
when the qdS approximation breaks down.}
\label{fig:num1}
\end{center}
\end{figure}

We first show two plots of the gravitational potential in figure \ref{fig:num1}. 
One generally expect that for larger wavenumber $k$, the natural
time variable $x$ in \Ref{xdef} is larger, so
if we were to series expand the Bessel (and Lommel) functions, we would need to keep more terms.
In other words, for large wavenumber $k$, the solutions ``explore'' the Bessel functions 
more, and the oscillations there can carry over to the gravitational potential. 

\subsection{Numerical solution: density contrast}
We form the density contrast
\be
\delta \equiv {\delta \rho - 3 H \delta u \over \rho}
\ee
where $\rho$ is the background matter density.
This is used to compute the growth factor. 
\begin{figure}
\begin{center}
\includegraphics[width=0.42\textwidth]{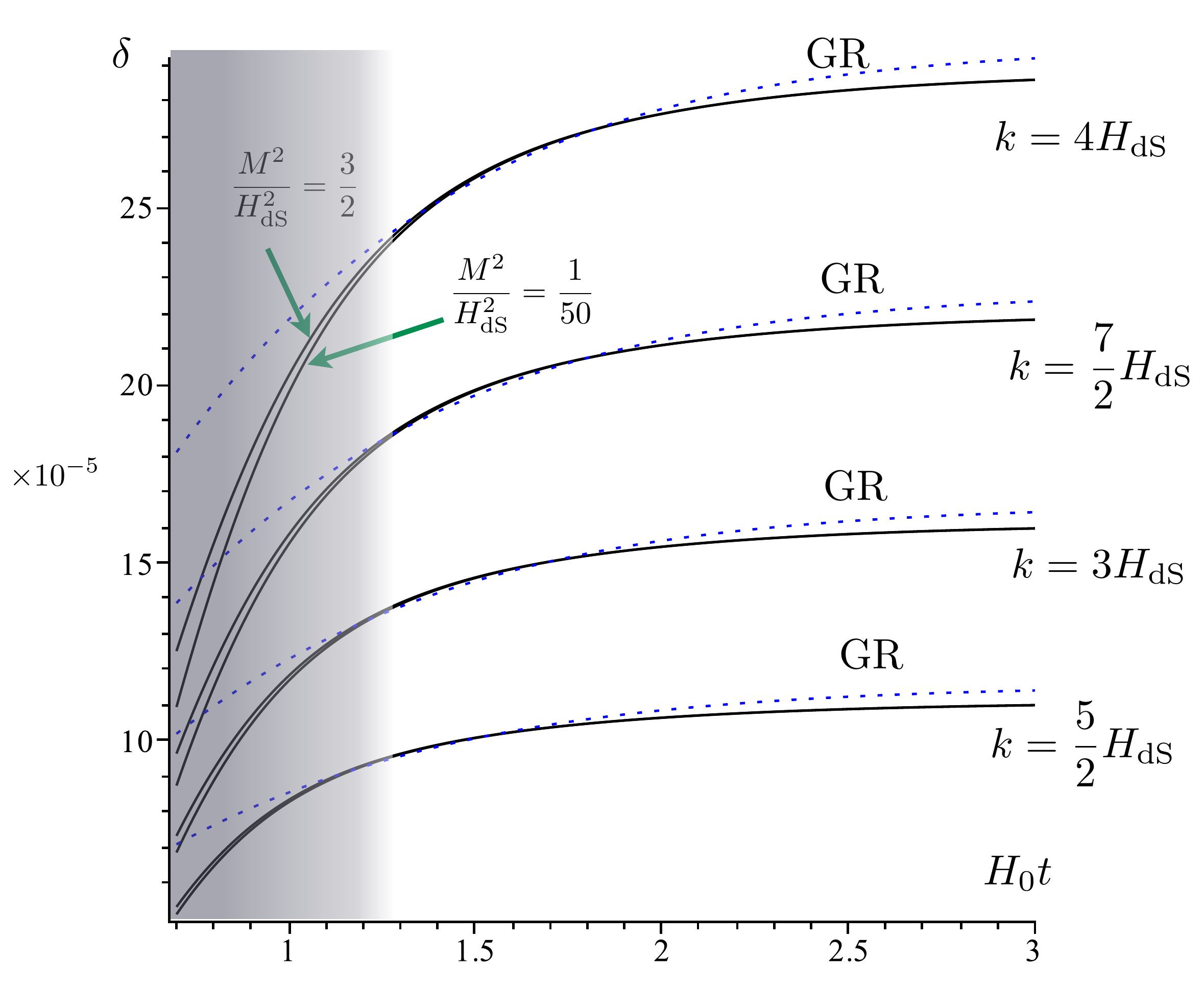} \hspace{2mm}
\includegraphics[width=0.42\textwidth]{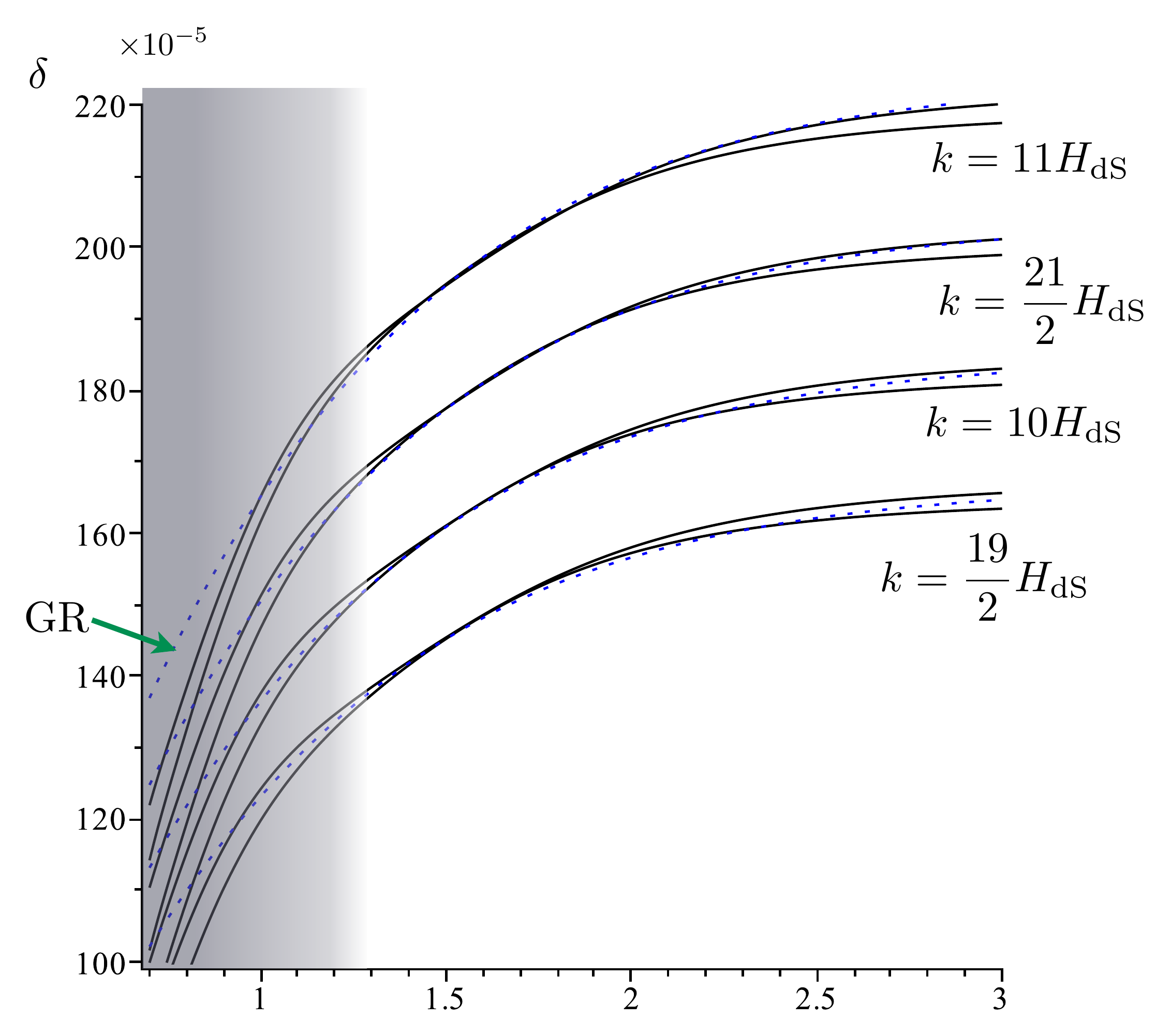}
\caption{Numerical qdS plots of the density contrast $\delta$
for various wavenumbers $k$.
Left panel:  small $k$ values, $k/H_{\rm dS}=
\{ 5/2, 3, 7/2, 4\} $. Right panel: intermediate $k$ values, $k/H_{\rm dS}=
\{ 19/2, 10, 21/2, 11\} $.
Each curve is plotted for two $M$ values, $M^2/H_{\rm dS}^2 = \{ 1/50,  3/2 \}$,
but the curves for the two $M$ values are nearly coincident in some cases.
The shaded area is our estimate for
when the qdS approximation breaks down. }
\label{fig:delta}
\end{center}
\end{figure}

We observe that although the $\Psi_+$ field
depends on $M$, the density contrast $\delta$ seems to depend on $M$ much less.
This is partially because of the way we fix boundary conditions,
which is imposed directly on $\delta$ and therefore only indirectly on $\Psi_+$. 
Nevertheless, although
the $\delta$ we see here does not differ appreciably from GR
around $H_0 t \gtrsim 1.3$, it does differ in the future,
so we would expect that if we go away some time
from the point at which we give the ICs (here $H_0 t = 1.5$), 
there would in fact be some controllable difference,
which is where the phenomenology of matter perturbations could begin.

It is of great importance whether there is a Vainshtein-like mecanism  here.
If there is a finite gap between the GR solution and the bimetric solution
for {\it any} value of $M$, one might be tempted to conclude that
there is in fact such a mechanism at work. However, there are also other parameters,
for example $c_J$, that can be turned on to try to mimic GR at zeroth order.
See figure \ref{deltacj}.

\subsection{Numerical solution: massive wave}

\begin{figure}
\begin{center}
\includegraphics[width=0.4\textwidth]{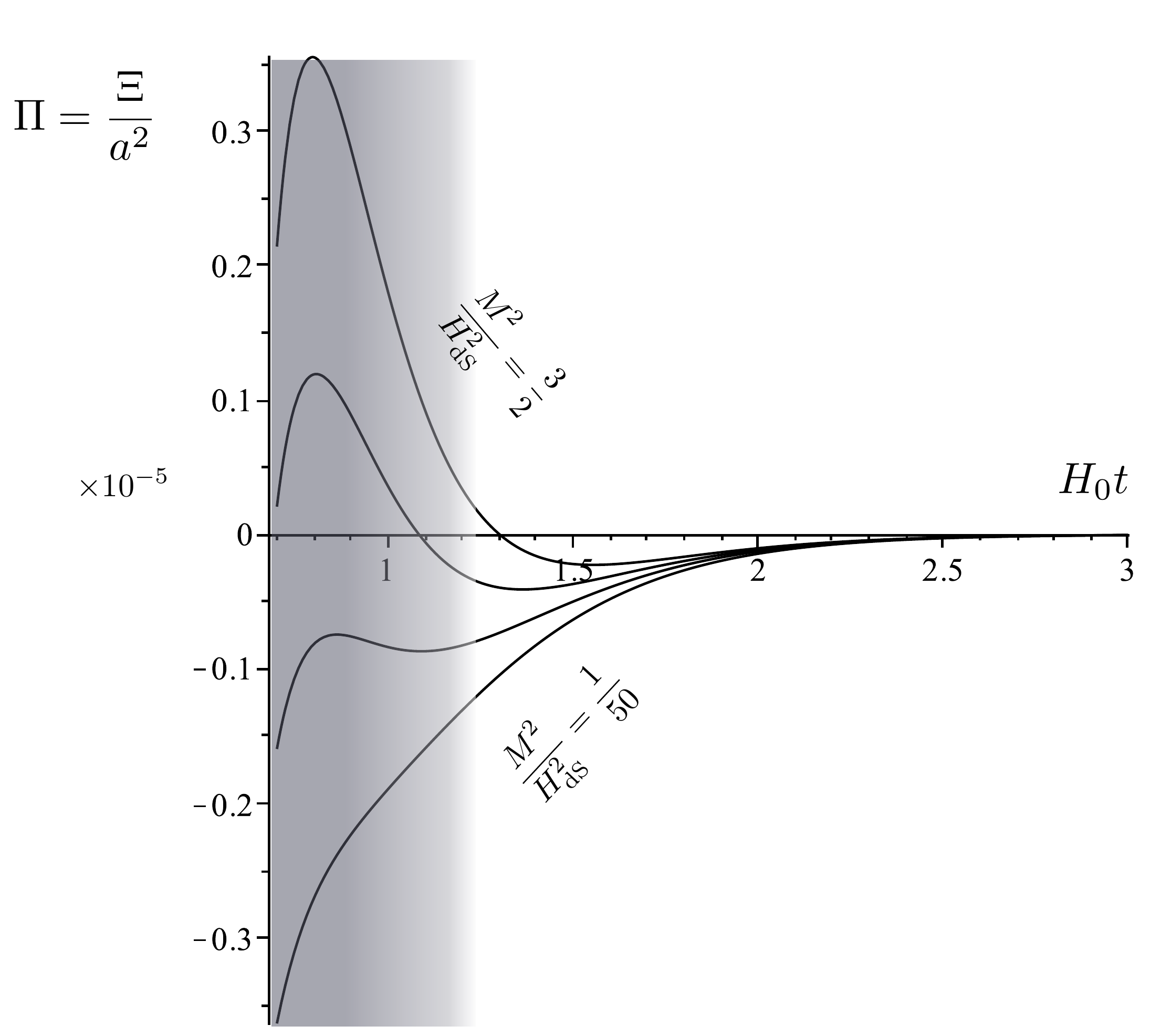} 
\includegraphics[width=0.45\textwidth]{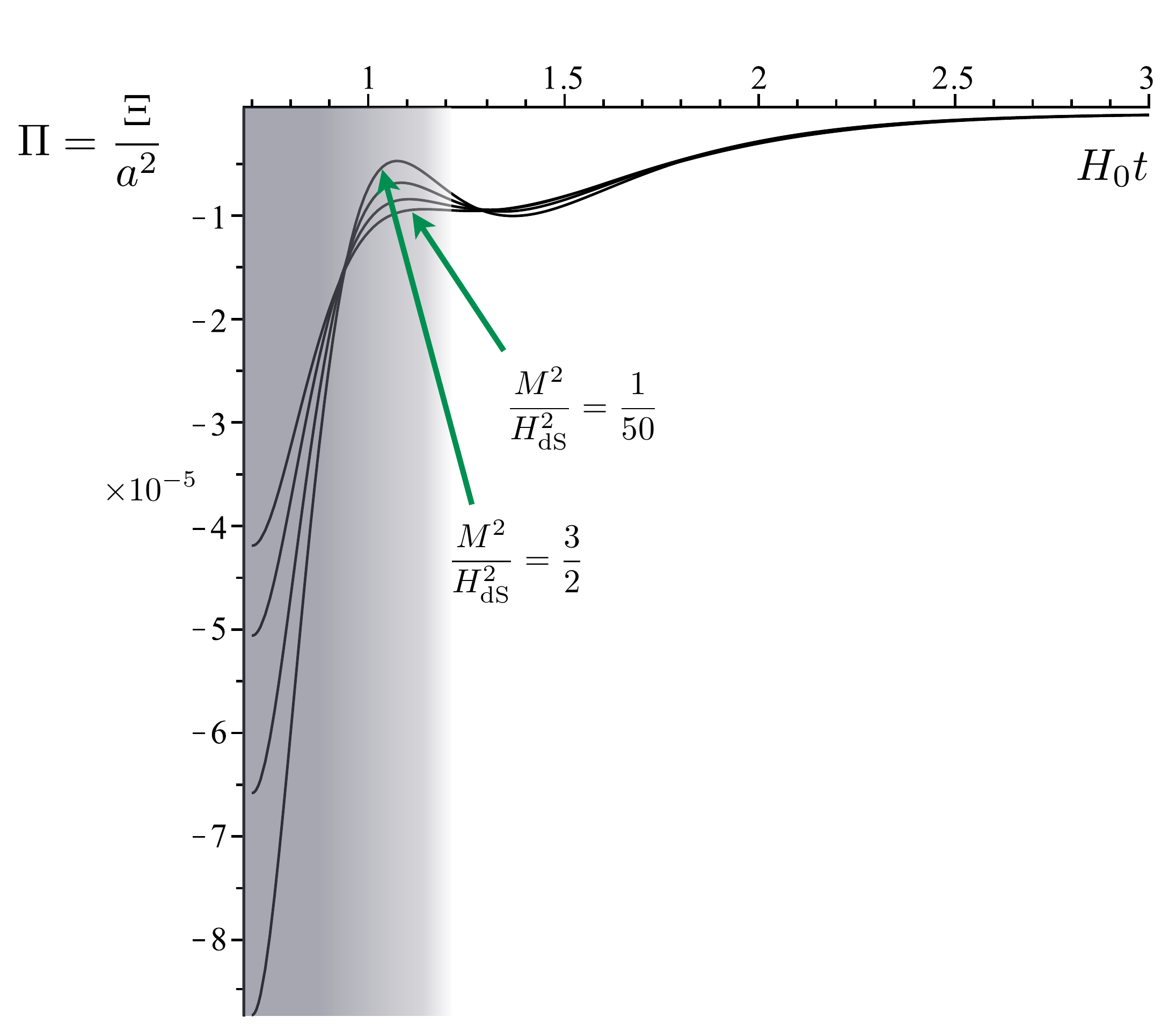} 
\caption{Numerical qdS plots for the massive scalar
gravitational fluctuation
$\Pi=\Xi/a^2$, for wavenumber $k=(5/2) H_{\rm dS}$ (left panel) 
and $k=10  H_{\rm dS}$ (right panel), and 
for mass parameter $M^2/H_{\rm dS}^2 = \{ 1/50, 4/5, 5/4, 3/2 \}$.}
\label{Xifig}
\end{center}
\end{figure}

The massive wave $\Pi$/$\Xi$  is  plotted in Fig. \ref{Xifig}.
Again, the ``bumps'' are in the region where the 
approximation has already broken down, and so cannot be trusted.
Still, also here we learn something about 
the massive wave around $H_0 t \sim 1.5$, 
and we see that it does depend on $M$, as one would expect.

\begin{figure}
\begin{center}
\includegraphics[width=0.4\textwidth]{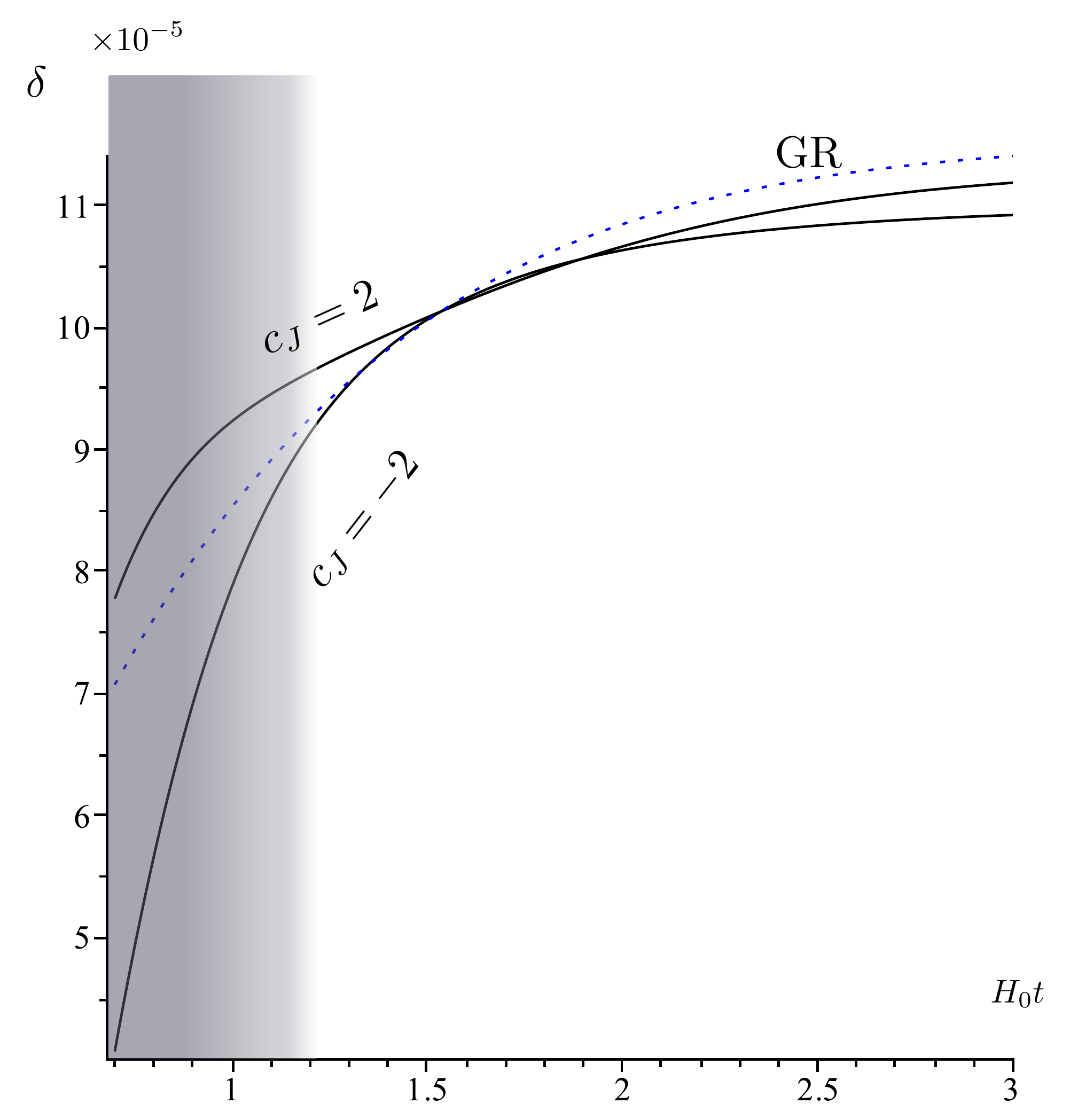} 
\caption{Numerical qdS plots for the density contrast $\delta$, for 
the homogeneous mode in $\Xi$ turned on, i.e.\ 
$c_J\neq 0$, unlike in the other plots in this section,
where $c_J=0$. }
\label{deltacj}
\end{center}
\end{figure}

One can use the existence of the homogenous
mode to see if one can recreate GR. 
We show some simple attempts in this direction in figure \ref{deltacj}.

\section{Towards $\Lambda$CDM}
We see that it is in principle possible to obtain good accuracy
with the quasi-de Sitter approximation,
but what suffices depends on the detailed application.
In particular, with our current understanding it seems that it would  be beneficial
to automate an arbitrary-order qdS approach in symbolic manipulation software,
especially if one wants to go to relevent eras such as $z \sim 1$. 
This is certainly possible, but outside the scope of this work.
Even better would be if one could
solve the fluctuation equations in the exact solution. At the moment
we do not know if this is feasible in practice.

\section{Conclusions and outlook}

We have computed the scalar fluctuation equations in de Sitter space
and quasi de Sitter space, and we have found some analytical
and some numerical solutions. There is much left to do as regards phenomenology. 

It would also be very interesting to perform the analogous Hamiltonian 
analysis to analyze linear and nonlinear stability of these fluctuations.
In GR, we can solve the gravitational perturbation separately,
which then completely determine the matter perturbations.
It is not clear that this strategy should automatically work here, as some of the constraints may become dynamical.
But at least in quasi-de Sitter, there seems to be no real issue with this.  
To completely understand this issue, we would also need to perform a Hamiltonian analysis,
which is beyond the scope of this work.

\section{Acknowledgements}

We thank the entire ``Dark Energy Working Group'' 
at the Oskar Klein Center for useful discussions.
We also thank Gregory Gabadadze, Claudia de Rham and Claes Uggla
for useful discussions in earlier stages of this work. 
M.B. and E.M. thank the Swedish Vetenskapsr{\aa}det for financial support.

\appendix

\section{The inhomogenous Bessel equation } 
\label{app:bessel}

\subsection{General}
Let us 
begin by recalling some elementary facts about
general inhomogenous 2nd order ODEs to set notation:
\be \label{ODE}
D[y] = y'' + p(x) y' + q(x) y = g(x) \; . 
\ee
for some polynomials $p(x)$, $q(x)$. 
Let $y_1$ and $y_2$ be fundamental
(linearly independent and normalized)  solutions
to the {\it homogenous} ODE  $D[y]=0$. 
We set for the general solution
\be \label{yg}
y_{\rm g}(x) = u_1(x)y_1(x)+u_2(x)y_2(x) \; ,  
\ee
for two unknown coefficient functions $u_1$, $u_2$.
Using variation of parameters we 
 find solutions for $u_1(x)$ and $u_2(x)$:
\be
u_1(x) = -\int \!dx\,  {y_2(x) g(x) \over W(x)}   \; , \quad u_2(x) = \int \!dx\,  { y_1(x) g(x)  \over W(x)} \; . 
\ee
where the Wronskian $W$ is the usual determinant
\be
W[y_1,y_2] = y_1 y_2' - y_2 y_1' \; . 
\ee

\subsection{Bessel}
For Bessel functions, the
fundamental solutions are
$y_1(x)=J_{\nu}(x)$ and  $y_2(x)=Y_{\nu}(x)$
and the  Wronskian is quite simple:
\be
W[J_{\nu},Y_{\nu}] = J_{\nu+1}Y_{\nu}-J_{\nu}Y_{\nu+1} = {2 \over \pi x}
\ee
so the coefficient functions become
\be
u_1(x) = -{\pi \over 2}\int  \!dx\,  Y_{\nu}(x) g(x) \cdot x   \; , \quad u_2(x) = {\pi \over 2}\int  \!dx\,  J_{\nu}(x) g(x)  \cdot x \; . 
\ee
where again $g(x)$ is the right-hand side of the inhomogenous equation, 
and the general (and generic) solution is simply (\ref{yg}):
\be  
y_{\rm g} &=&  u_1(x)y_1(x)+u_2(x)y_2(x) \\
&=& \left(-{\pi \over 2}\int  \!dx\,  Y_{\nu}(x) g(x) \cdot x \right)J_{\nu}(x)
+ \left({\pi \over 2}\int  \!dx\,  J_{\nu}(x) g(x)  \cdot x   \right)Y_{\nu}(x)    \; . \label{generalsol}
\ee
A simple way to specify ICs is to fix $\int_{x_0}^x$ in (\ref{generalsol}) and write
\be
y = y_{\rm h} + y_{\rm g} \; ,
\ee
with the usual two free integration constants in
the homogeneous piece $y_{\rm h}$, and no
free constants in $y_{\rm g}$. 
Then (\ref{generalsol}) above represents the {\it particular} solution.
If we specialize to a power-like right-hand side, $g(x)=x^{\mu}$, 
and fix $x_0=0$, then
\be  \label{genericpart}
y_{\rm particular}(x) & =& \left(-{\pi \over 2}\int_0^x  \!dx\,  Y_{\nu}(x) x^{\mu+1}  \right)J_{\nu}(x)
+ \left({\pi \over 2}\int _0^x \!dx\,  J_{\nu}(x) x^{\mu+1}     \right)Y_{\nu}(x) \\
&=&s_{\mu+1,\nu}(x)
\ee
where $s_{\mu,\nu}(x)$ is a Lommel function. 
The series expansion of this Lommel function at $x=0$ is
\be
s_{\mu,\nu}(x)  = {x^{\mu+1} \over (\mu-\nu+1)(\mu+\nu+1)} + \O(x^{\mu+3})
\ee
i.e. the order of vanishing at $x=0$ is independent of $\nu$,
which is not evident from the integral representation \Ref{genericpart}. 

\section{Massless limit}
In the limit $M\rightarrow 0$ ($\beta_2\rightarrow 0$)
we find 
\be
J_{3/2}(x) = \sqrt{2 \over \pi}\, {\sin x -  x \cos x \over x^{3/2}}\; , \quad
Y_{3/2}(x) = -\sqrt{2 \over \pi}\, {x \sin x +  \cos x \over x^{3/2}}
\ee
Using this, we find from \Ref{genericpart}
that the associated Lommel functions reduce to simple powers:
\be
s_{3/2,3/2}(x) &=& {x^2+2 \over x^{3/2}} \\
s_{5/2,3/2}(x)&=& x^{3/2} \; . 
\ee

\section{Replacement rules}
\label{app:repl}

The range of $t$ is such that the dimensionless $\epsilon(t)$ is considered small. In the linear $\epsilon$   expansion, we have for the scale factors 
\be
a(t) &=& a_{\rm dS}(1-a_{\epsilon}\epsilon)\\
Y(t) &=& Y_{\rm dS}(1-a_{\epsilon} c_Y \epsilon)
\ee
with
\be
a_{\epsilon}&=& {1 \over 9} \\
c_Y &=& {M^2 + 4(1+c^2)H_{\rm dS}^2 \over M^2-2(1+c^2)H_{\rm dS}^2}  \; ,
\ee
and for the derived quantities
\be
H(t) &\equiv & {\dot{a} \over a} = H_{\rm dS}(1 + H_{\epsilon}\epsilon(t)) \\
K(t) &\equiv &   {\dot{Y} \over Y} = H_{\rm dS}(1+ K_{\epsilon}\epsilon(t)) \\
X(t) &\equiv& {\dot{Y} \over \dot{a}} = c(1 + X_{\epsilon}\epsilon(t) )
\ee
where the coefficients are constants
\be
H_{\epsilon}&=& 3 a_{\epsilon}Y_{\epsilon} \\
K_{\epsilon} &=& {c_Y / 3} \\
X_{\epsilon} &=& 2a_{\epsilon}(Y_{\epsilon}-1) \; . 
\ee
These simple expressions are sufficient
to eliminate all derivatives on the background. 

\section{Quasi-de-Sitter expansion in Einstein gravity}
\label{app:GR}

In this appendix we apply the qdS expansion to scalar perturbations
in Einstein gravity. We compare the results of first and second
order qdS expansions to the exact
solution for the  two-component fluid
(with only dust and dark energy, which
should be a good approximation to physical cosmology in this time interval). 

The well known GR scalar perturbation equations for dust with the
conventions used in this paper can be written as:
\be
-\frac{1}{a^{2}}\nabla^{2}\Psi+3H\left(H\Phi+\dot{\Psi}\right)=-\frac{\delta\rho}
{2M_P^2}
\label{eq00}
\ee
\be
-\partial_{i}\left(\dot{\Psi}+H\Phi\right)=\frac{\partial_i \delta u}{2M^{2}_P}
\label{eq0i}
\ee
\[
\ddot{\Psi}+H\dot{\Phi}+3H\left(H\Phi+\dot{\Psi}\right)+2\dot{H}\Phi=0
\]
\[
-\frac{1}{2a^{2}}\partial^i\partial_j\left(\Phi-\Psi\right)=0
\]
Recall from eq. \Ref{qexp} that the scale factor expanded to second order
can be written as:
\[a(t) =a_{\rm dS}(t)\left( 1-{1 \over 9} \epsilon(t) -{1 \over 324}\epsilon(t)^2 \right)
\]
Using this expansion, the equations of motion for the perturbations can  easily be recast
in the general form
\[
D_{ij}\phi^{j}=\left(D_{ij}^{0}+D_{ij}^{\varepsilon}+D_{ij}^{\varepsilon^{2}}\right)\left(\phi_{0}^{j}+\phi_{\varepsilon}^{j}+\phi_{\varepsilon^{2}}^{j}\right)=\mathcal{J}_{i}^{0}+\mathcal{J}_{i}^{\varepsilon}+\mathcal{J}_{i}^{\varepsilon^{2}}
\]
which we split order by order into three sets of equations:
\[
D_{ij}^{0}\phi_{0}^{j}=\mathcal{J}_{i}^{0}
\]
\[
D_{ij}^{0}\phi_{\varepsilon}^{j}=\mathcal{J}_{i}^{\varepsilon}-D_{ij}^{\varepsilon}\phi_{0}^{j}
\]
\[ 
D_{ij}^{0}\phi_{\varepsilon^{2}}^{j}=\mathcal{J}_{i}^{\varepsilon^{2}}-D_{ij}^{\varepsilon}\phi_{\varepsilon}^{j}-D_{ij}^{\varepsilon^{2}}\phi_{0}^{j}
\]
It is possible to analytically solve these equations, obtaining for
the gravitational potential:
\[
\Psi_{0}=C_{1}e^{-H_{\rm dS}t}+C_{2}e^{-3H_{\rm dS}t}
\]
\[
\Psi_{\varepsilon}=\frac{8}{15}\left(5C_{1}e^{2H_{\rm dS}t}+3C_{2}\right)e^{-6H_{\rm dS}t}
\]
\[
\Psi_{\varepsilon^{2}}=\frac{7}{90}\left(50C_{1}e^{2H_{\rm dS}t}+27C_{2}\right)e^{-9H_{\rm dS}t}
\]
where we have set the integration constants of the 1st and 2nd order equations to zero.
In fact, the solutions of the homogenous versions of
these equations are effectively of zeroth order,
so to be consistent with the $ $qdS expansion we should turn them off.
With this understanding, the three expressions above are nicely separated in order. In the language of the main text, we have $\Delta_{\Psi_0} = 1$, $\Delta_{\Psi_{\epsilon}}=4$,
$\Delta_{\Psi_{\epsilon^2}}=7$ (which is a compact way of stating that the leading terms
for large $t$ are $e^{-H_{\rm dS}t}$, $e^{-4H_{\rm dS}t}$ and $e^{-7H_{\rm dS}t}$, respectively) . Here the suppressed terms in each fluctuation 
are less suppressed than the next order, as one would expect. This will be different for the matter perturbations below. 

It is then straightforward to compute energy density and velocity
perturbations using \Ref{eq00} and \Ref{eq0i}. For the energy density we obtain:
\[
{\delta\rho_{0} \over M_P^2} =\left(12C_{2}H_{\rm dS}^{2}-\frac{2k^{2}}{c_2^2}C_{1}\right)e^{-3H_{\rm dS}t}-\frac{2k^{2}}{c_2^2}C_{2}e^{-5H_{\rm dS}t}
\]
\[
{\delta\rho_{\varepsilon}\over M_P^2}=\frac{4}{15}\left(135H_{\rm dS}^{2}C_{1}e^{2H_{\rm dS}t}-\frac{30k^{2}}{c_2^2}C_{1}+225H_{\rm dS}^{2}C_{2}-\frac{22k^{2}}{c_2^2}C_{2}e^{-2H_{\rm dS}t}\right)e^{-6H_{\rm dS}t}
\]
\[
{\delta\rho_{\varepsilon^{2}}\over M_P^2} =\frac{1}{45}\left(7560H_{\rm dS}^{2}C_{1}e^{2H_{\rm dS}t}-\frac{810k^{2}}{c_2^2}C_{1}+7452 H_{\rm dS}^{2}C_{2}-\frac{521k^{2}}{c_2^2}C_{2}e^{-2H_{\rm dS}t}\right)e^{-9H_{\rm dS}t} \; . 
\]
We observe that
$\Delta_{\delta \rho_0} = 3$, $\Delta_{\delta\rho_{\epsilon}}=4$,
$\Delta_{\delta\rho_{\epsilon^2}}=7$. 0
Because the fields begin to mix at first order in the qdS approximations,
also fields that are suppressed at zeroth order, as $\delta\rho$ is,
receive a first correction that is relatively big if the other fields
are relatively big. In particular, $\delta\rho_0$ starts at  $e^{-3 H_{\rm dS}t}$
and the linear qdS field has a $e^{-4 H_{\rm dS}t}$ piece.
Moreover, there can be terms in the lower order fields,
just from solving the equations, that strictly belong to higher orders in the expansion;
this is the case for the $e^{-5 H_{\rm dS}t}$ term in $\delta\rho_0$.
We will typically keep such terms at the order at which they appear, but they cannot be considered reliable
for truncation at the given order. 

For the velocity perturbation:
\[
{\delta u_{0}\over M_P^2}=4C_{2}H_{\rm dS}e^{-3H_{\rm dS}t}
\]
\[
{\delta u_{\epsilon}\over M_P^2}=12H_{\rm dS}\left(C_{1}e^{-4H_{\rm dS}t}+C_{2}e^{-6H_{\rm dS}t}\right)
\]
\[
{\delta u_{\epsilon^{2}}\over M_P^2}=\frac{4}{5}H_{\rm dS}\left(40C_{1}e^{-7H_{\rm dS}t}+29C_{2}e^{-9H_{\rm dS}t}\right)
\]
for which $\Delta_{\delta u_0}=3$,  $\Delta_{\delta u_{\epsilon}}=4$
and $\Delta_{\delta u_{\epsilon^2}}=7$, like for $\delta\rho$. 

To compare these results with the $\Lambda$CDM solutions, in the
following called $\Psi_{f}$, $\delta\rho_{f}$, and $\delta u_{f}$,
we have chosen $C_{1}$ and $C_{2}$ such that $ $$\Psi_{f}$ asymptotically matches
$\Psi_{0}$  in the future. Notice that this is a slightly different choice of 
integration constants from that used for bimetric theory in the main text, 
but it is more suited to this analysis. Typically the two choices give very similar results. 

\begin{figure}[h]
\begin{centering}
\includegraphics[scale=0.27]{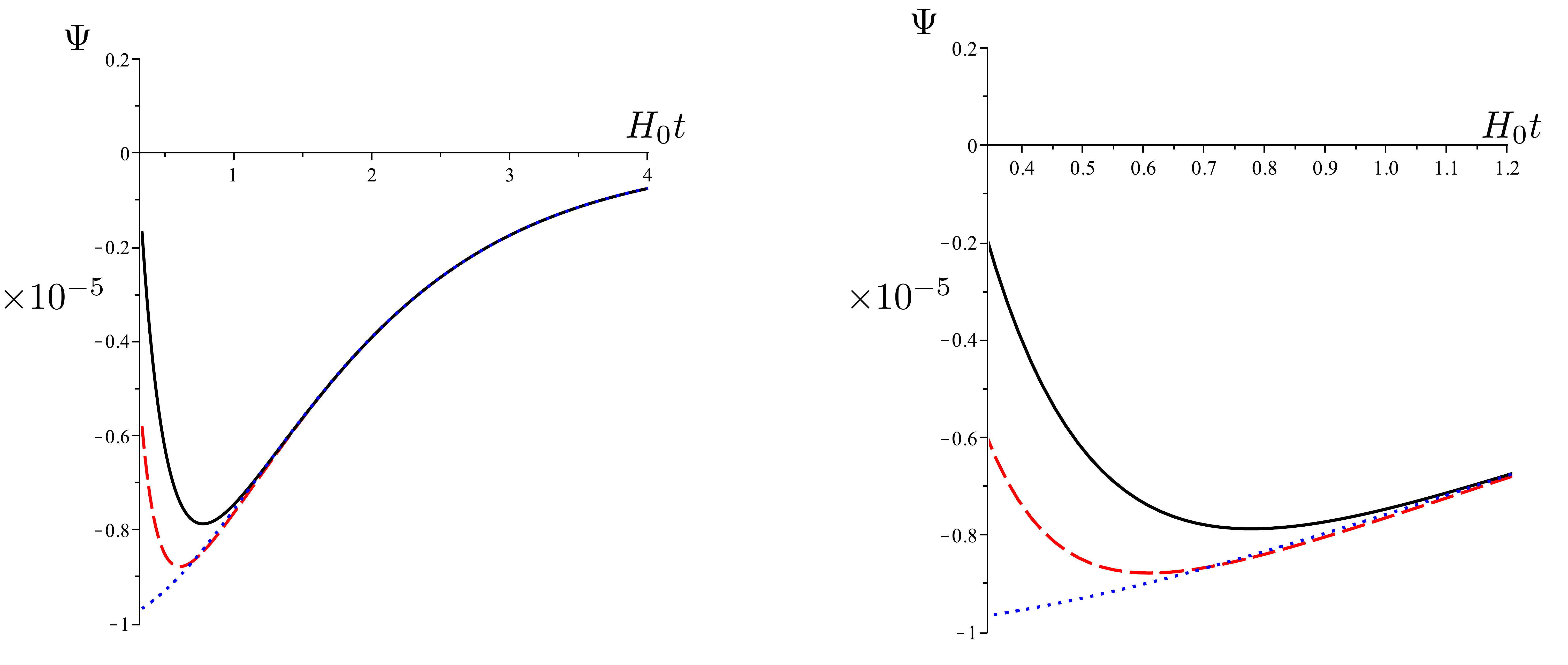}
\par\end{centering}
\caption{Comparing approximations for the GR gravitational potential $\Psi$. The first order qdS approximation is black solid,
the second order qdS approximation is red dashed, and the  
exact $\Lambda$CDM  solution is blue dotted.}
\label{Gr_psi}
\end{figure}

With the initial conditions $\Psi_{f}({t}_{\star})=10^{-5}$, $\dot{\Psi}_{f}({t}_{\star})=0$ ,
where ${t}_{\star}$ is $H_0 {t}_{\star}=0.002$ (during recombination)
we found\footnote{We have not been careful 
with the overall factor here, since we do no actual phenomenology in this paper.
If the factor changes, all fields would simple be multiplied by the same correction factor,
since we are doing linear perturbation theory.}
\be
C_{1}\simeq-2.28 \cdot 10^{-5}\; , \quad C_{2}\simeq4.89 \cdot 10^{-5} \; . 
\ee
Perhaps the best way to get an intuitive idea about how good our first and second order qdS approximations are is to consider $\Psi$ plots like those of figure \ref{Gr_psi}.
A more precise measure is the relative error:
\[
\left|\frac{\Psi_f-\left(\Psi_0+\Psi_\epsilon \right)}{\Psi_f} \right|_{H_{0}t=1}\simeq 0.015\; , \quad
\left|\frac{\Psi_f-\left(\Psi_0+\Psi_\epsilon+\Psi_{\epsilon^2} \right)}{\Psi_f} \right|_{H_{0}t=1}\simeq 0.009 \; . 
\]
so both the first and second order expansions are good to about 1\% around present. For our purposes, it is also important to have an idea when the approximations break down. We find that
\[
\left|\frac{\Psi_f-\left(\Psi_0+\Psi_\epsilon \right)}{\Psi_f} \right|_{H_{0}t=0.7}\simeq 0.1\,\,\,\,\,\,\,\,\,\,\,\, 
\left|\frac{\Psi_f-\left(\Psi_0+\Psi_\epsilon+\Psi_{\epsilon^2} \right)}{\Psi_f} \right|_{H_{0}t=0.5}\simeq 0.1
\]
i.e.\ we are down to 10\% accuracy at $H_0 t \sim 0.7$ and $H_0 t \sim 0.5$
for the first and second order qdS approximations, respectively.

\begin{figure}[h]
\begin{centering}
\includegraphics[scale=0.37]{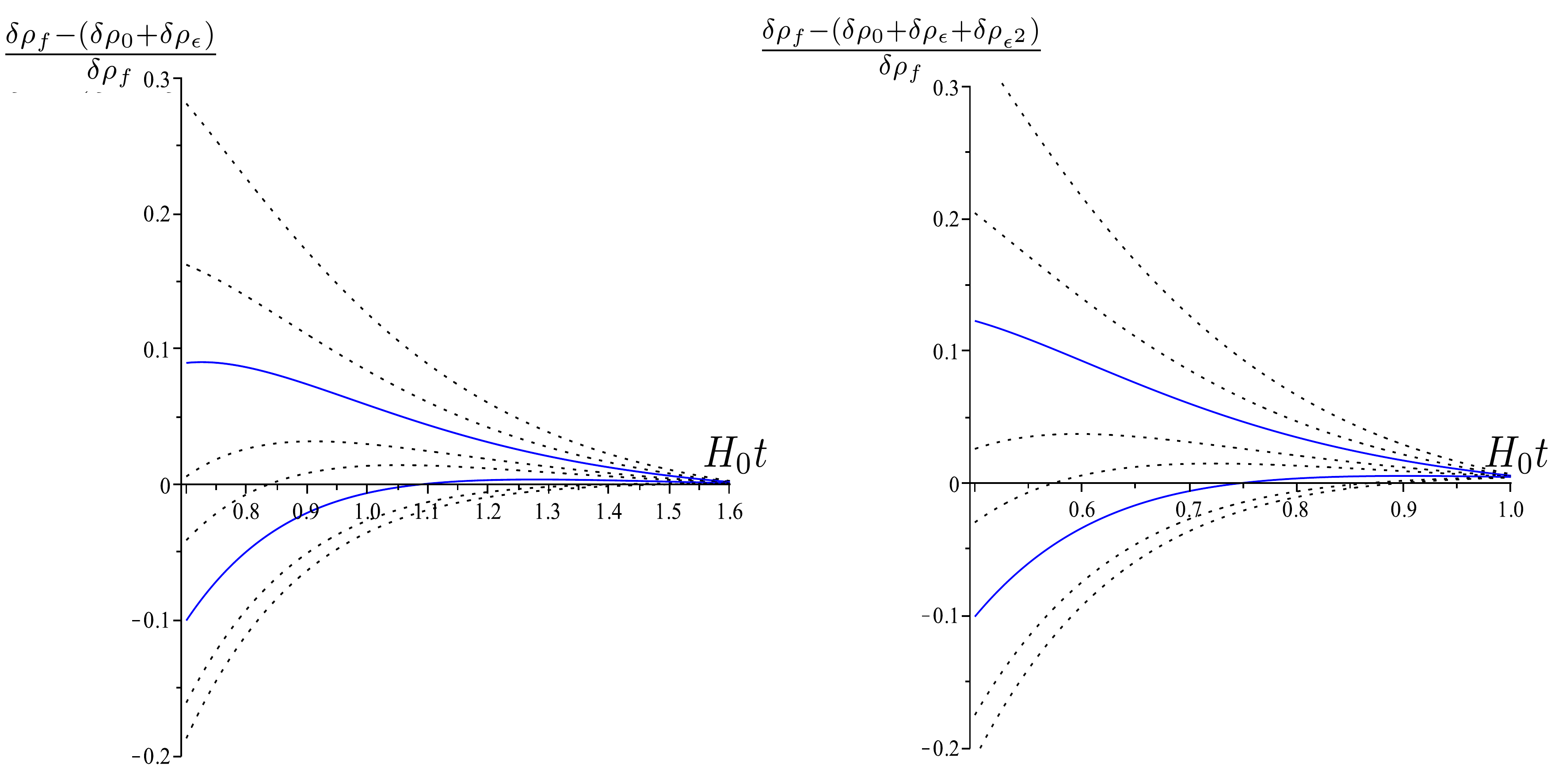}
\par\end{centering}
\caption{Signed percentage differences between density perturbation expansions and exact solutions for various $k$ values. Left panel: differences between 1st order qdS and exact solutions. Right panel: differences between 2nd order qdS and exact solutions. In the interval $3H_{\rm dS} \lesssim k^{2} \lesssim 12H_{\rm dS}$ (solid blue lines) 
the percentage differences are always smaller than 10\% back to $H_{0}t\simeq0.7$ in the case of 1st order qdS, and back  to $H_{0}t\simeq0.5$ in the case of 2nd order qdS. }
\label{rho_k}
\end{figure}

\begin{figure}[h]
\begin{centering}
\includegraphics[scale=0.25]{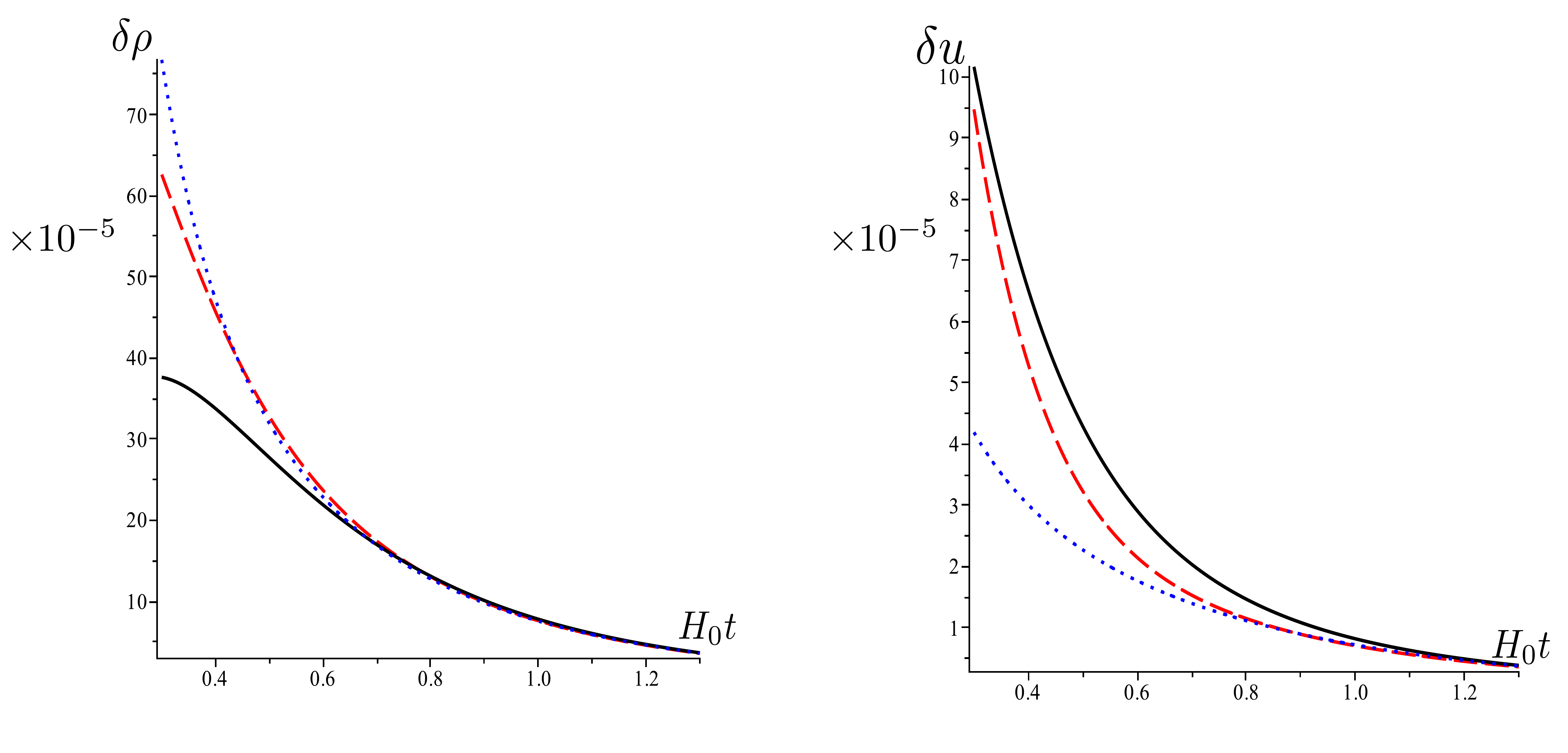}
\par\end{centering}
\caption{ Density perturbations for $k^2=5H_{\rm dS}$ and velocity perturbations:   the first order qdS approximation is black solid,
 the second order qdS approximation is red dashed, and the exact $\Lambda CDM$  solution is blue dotted}
\label{rho}
\end{figure}

The range of validity of the various approximations for the matter perturbations might vary  with $k$. We analyzed what happens if $1/k^2$ lies between the horizon scale and two orders of magnitude below the horizon scale, i.e. when:
\[
H_{\rm dS}^2<k^2<100H_{\rm dS}^2
\]
Around present, for all $k$ in this range,  the approximations do not differ more than 10\% from the exact solution, i.e.
\[
0.1 \lesssim \left|\frac{\delta\rho_f-\left(\delta\rho_0+\delta\rho_\epsilon \right)}{\delta\rho_f} \right|_{H_{0}t=1}\lesssim 0.01\; , \quad
0.007\lesssim \left|\frac{\delta\rho_f-\left(\delta\rho_0+\delta\rho_\epsilon+\delta\rho_{\epsilon^2} \right)}{\delta\rho_f} \right|_{H_{0}t=1}\lesssim 0.005 \; . 
\]
To be more precise, we found a small range of $k$ in wich the density perturbation qdS expansions are as good as the $\Psi$ expansions or even better, see figure \ref{rho_k} and \ref{rho}, in particular we find:
\[
\begin{cases}
\begin{array}{c}
3H_{\rm dS} \lesssim k^{2} \lesssim 12H_{\rm dS}\\
H_{0}t > 0.7
\end{array}  &   \,\,\,\Longrightarrow\,\,\,\,\,\,\,\, \left|\frac{\delta\rho_f-\left(\delta\rho_0+\delta\rho_\epsilon \right)}{\delta\rho_f} \right |< 0.1     \end{cases}
\]
\[
\begin{cases}
\begin{array}{c}
3H_{\rm dS} \lesssim k^{2} \lesssim 12H_{\rm dS}\\
H_{0}t > 0.5
\end{array}  &   \,\,\,\Longrightarrow\,\,\,\,\,\,\,\, \left|\frac{\delta\rho_f-\left(\delta\rho_0+\delta\rho_\epsilon +\delta\rho_{\epsilon^2} \right)}{\delta\rho_f} \right|< 0.1     \end{cases}
\]

Having computed density and velocity perturbations qdS expansions, it is simple to compute the corresponding expansions for the comoving density contrast $\delta = (\delta \rho - 3 H \delta u)/\rho$:
\[
\delta_1= \delta_{0}+ \delta_{\epsilon} \,\,\,\,\,\,\,\,\,\,\,\,\,\,\,\,\, \delta_2= \delta_{0}+ \delta_{\epsilon}+ \delta_{\epsilon^2}
\]
%\[
%\delta_1\simeq \frac{\left(\delta\rho_0+ \delta\rho_\epsilon \right)-3H\left(\delta u_0+ \delta u_\epsilon \right)}{\rho} \,\,\,\,\,\,\,\,\,\,\,\,\,\,\,\,\delta_2\simeq =\frac{\left(\delta\rho_0+ \delta\rho_\epsilon + \delta\rho_{\epsilon^2} \right)-3H\left(\delta u_0+ \delta u_\epsilon +\delta u_{\epsilon^2} \right)}{\rho}
%\]
Concerning the validity of $\delta$ approximations
we find the same results as for $\delta\rho$ above. This is simply because the errors in the $\delta u$ expansion are  smaller than the errors
in the $\delta\rho$ expansions for all values of $k^2$.

Concerning the growth index we find  that in general, the values computed with our 1st and 2nd order qdS approximations are trustable only for  $H_0 t\gtrsim1.5$
It is not hard to understad why this happens. We recall the explicit expression
for the growth index:
\[
\gamma=\frac{\ln f}{\ln\left( \Omega_m/a^3 \right)} \; ,
\]
with
\[
f =   {d \ln \delta \over d \ln a }  .
\]
Now, given errors 
in $\delta$ and $\dot{\delta}$, it is straightforward to compute the expected error in $\gamma$:
\[
\Delta\gamma=\left( -\frac{\Delta \delta}{\delta} +\frac{\Delta \dot{\delta}}{\dot{\delta}} \right) \frac{1}{\ln\left( \Omega_m/a^3 \right)}
\]
In general for $H_0 t\lesssim1.5$ we have  $|\Delta\dot{\delta}|\gg |\Delta{\delta}|$. This is why the growth index expansions start deviating from the exact function before (in the sense of coming from the future) the other quantities of interest, i.e.\ the approximation for $\gamma$ itself is a little worse
than that for e.g. $\Psi$.

\begin{figure}[h]
\begin{centering}
\includegraphics[scale=0.27]{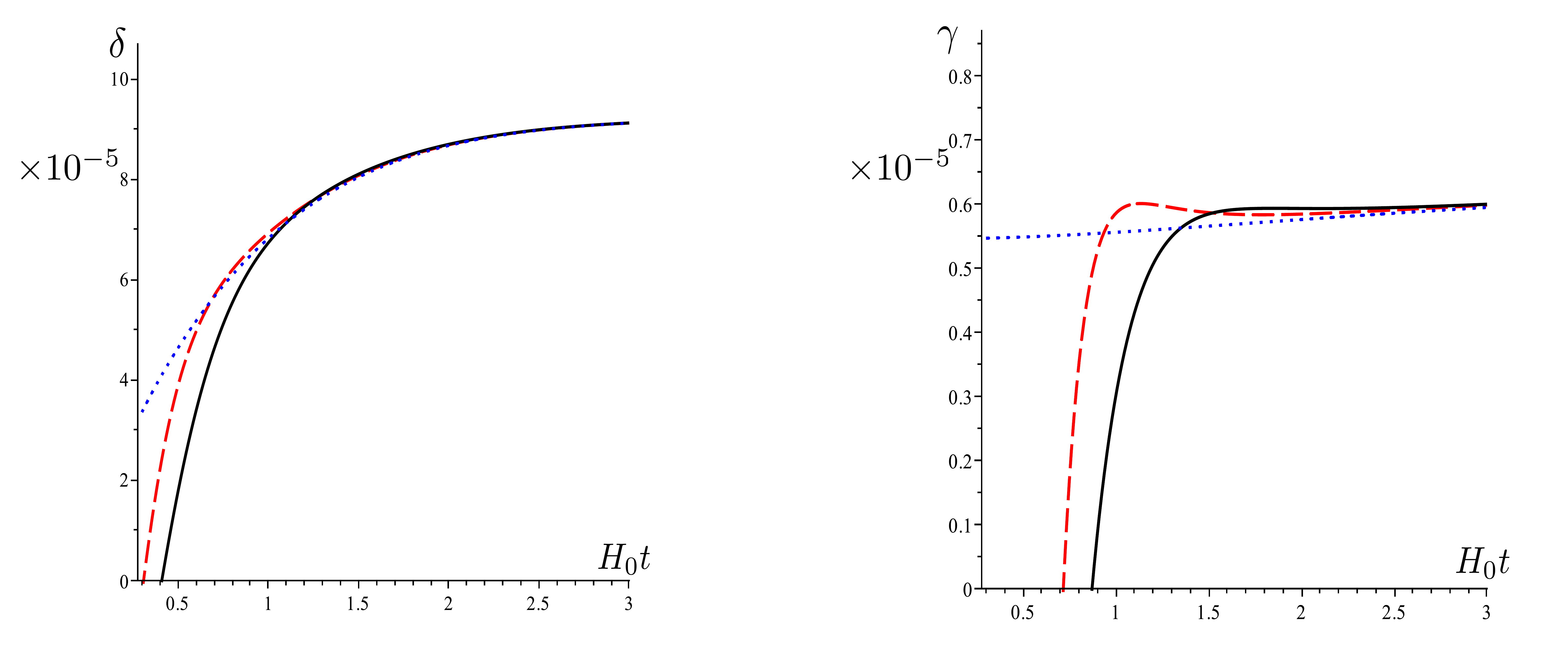}
\par\end{centering}
\caption{ Comoving density contrast $\delta$ and growth index 
$\gamma$ for $k^2=5H_{\rm dS}$:  the second order qdS approximation, dashed,  the first order, solid, and the $\Lambda$ CDM  solution, dotted.}
\label{Gr_delta_gamma}
\end{figure}

\section{Bimetric tensor modes in de Sitter}
\label{app:tensor}

We are not directly interested in tensor models in this paper,
but we would like to compare our mass parameters
to the mass parameter of the tensor fluctuation.
Consider gravitational waves traveling in the $z$-direction. The conditions for tracelessness and divergence-freeness of the perturbations are solved by the following ansatz 
\begin{align}
 \dd s_g^2 & = - \dd t^2 + a(t)^2 \left(\dd \vec x^2 + 2h^g_{xy}(t,z) \dd x \dd y + h^g_{xx}(t,z) (\dd x^2-\dd y^2) \right) \,, \\
 \dd s_f^2 & = - c^2 \dd t^2 + c^2 a(t)^2 \left(\dd \vec x^2 + 2h^f_{xy}(t,z) \dd x \dd y + h^f_{xx}(t,z) (\dd x^2-\dd y^2) \right) \,,
\end{align}
where  $a(t) = c_2 e^{H_{\rm dS} t}$.
In terms of the linear combinations
\begin{align}
 h_{xx}^+ = h_{xx}^g + c^2 h_{xx}^f \,, \quad h_{xx}^- = h_{xx}^g - h_{xx}^f \,, \\
 h_{xy}^+ = h_{xy}^g + c^2 h_{xy}^f \,, \quad h_{xy}^- = h_{xy}^g - h_{xy}^f \,,
\end{align}
the linearized equations of motion become
\begin{align}
 \left( \frac{\partial^2}{\partial t^2} + 3 H_{\rm dS} \frac{\partial}{\partial t} - \frac{1}{a^2} \frac{\partial^2}{\partial z^2} + M^2 \right) h^{-}_{xx,xy} = 0 \,, \\
 \left( \frac{\partial^2}{\partial t^2} + 3 H_{\rm dS} \frac{\partial}{\partial t} - \frac{1}{a^2} \frac{\partial^2}{\partial z^2}\right) h^{+}_{xx,xy} = 0 \,,
\end{align}
where
\begin{equation}
\label{FPM}
 M^2 = \left(1 + \frac{1}{c^2} \right) m^2 \left( c \beta_1 + 2 c^2 \beta_2 + c^3 \beta_3 \right)\; ,
\end{equation}
which can be compared to \Ref{eq:effectiveparameters}, and
 where we used the background $g$ and $f$ equations to reexpress
 $\beta_0$ and $\beta_4$ as
\begin{align}
 \beta_0 &= 3 H^2/m^2 - 3 \beta_1 c - 3 \beta_2 c^2 - \beta_3 c^3  \\
 c^4\beta_4 &= 3 c^2 H^2/m^2 - \beta_1 c - 3 \beta_2 c^2 - 3 \beta_3 c^3 \,.
\end{align}
%:


\begin{thebibliography}{99}

\bibitem{Weinberg:1988cp}
  S.~Weinberg,
  ``The Cosmological Constant Problem,''
  Rev.\ Mod.\ Phys.\  {\bf 61} (1989) 1.
  %%CITATION = RMPHA,61,1;%%


\bibitem{Polchinski:2006gy}
  J.~Polchinski,
  ``The Cosmological Constant and the String Landscape,''
  hep-th/0603249.
  %%CITATION = HEP-TH/0603249;%%

\bibitem{Dvali:2007kt}
  G.~Dvali, S.~Hofmann and J.~Khoury,
  ``Degravitation of the cosmological constant and graviton width,''
  Phys.\ Rev.\ D {\bf 76} (2007) 084006
  [hep-th/0703027 [HEP-TH]].
  %%CITATION = HEP-TH/0703027;%%

\bibitem{Polyakov:2009nq}
  A.~M.~Polyakov,
  ``Decay of Vacuum Energy,''
  Nucl.\ Phys.\ B {\bf 834} (2010) 316
  [arXiv:0912.5503 [hep-th]].
  %%CITATION = ARXIV:0912.5503;%%

\bibitem{Sjors:2011iv}
  S.~Sjors and E.~Mortsell,
  ``Spherically Symmetric Solutions in Massive Gravity and Constraints from Galaxies,''
  arXiv:1111.5961 [gr-qc].
  %%CITATION = ARXIV:1111.5961;%%



\bibitem{Baumann:2009ds}
  D.~Baumann,
  ``TASI Lectures on Inflation,''
  arXiv:0907.5424 [hep-th].
  %%CITATION = ARXIV:0907.5424;%%
  
\bibitem{Fierz:1939ix}
  M.~Fierz and W.~Pauli,
  ``On relativistic wave equations for particles of arbitrary spin in an electromagnetic field,''
  Proc.\ Roy.\ Soc.\ Lond.\ A {\bf 173} (1939) 211.
  %%CITATION = PRSLA,A173,211;%%


\bibitem{Volkov:2011an}
  M.~S.~Volkov,
  ``Cosmological solutions with massive gravitons in the bigravity theory,''
  JHEP {\bf 1201} (2012) 035
  [arXiv:1110.6153 [hep-th]].
  %%CITATION = ARXIV:1110.6153;%%

\bibitem{vonStrauss:2011mq}
  M.~von Strauss, A.~Schmidt-May, J.~Enander, E.~Mortsell and S.~F.~Hassan,
  ``Cosmological Solutions in Bimetric Gravity and their Observational Tests,''
  arXiv:1111.1655 [gr-qc].
  %%CITATION = ARXIV:1111.1655;%%

\bibitem{Volkov:2012cf}
  M.~S.~Volkov,
  ``Exact self-accelerating cosmologies in the ghost-free bigravity and massive gravity,''
  arXiv:1205.5713 [hep-th].
  %%CITATION = ARXIV:1205.5713;%%


\bibitem{Comelli:2011zm}
  D.~Comelli, M.~Crisostomi, F.~Nesti and L.~Pilo,
  ``FRW Cosmology in Ghost Free Massive Gravity from Bigravity,''
  arXiv:1111.1983 [hep-th].
  %%CITATION = ARXIV:1111.1983;%%



\bibitem{Hassan:2011vm}
  S.~F.~Hassan and R.~A.~Rosen,
  ``On Non-Linear Actions for Massive Gravity,''
  JHEP {\bf 1107} (2011) 009
  [arXiv:1103.6055 [hep-th]].
  %%CITATION = ARXIV:1103.6055;%%


\bibitem{Salam:1976as}
  A.~Salam and J.~A.~Strathdee,
  ``A Class of Solutions for the Strong Gravity Equations,''
  Phys.\ Rev.\ D {\bf 16} (1977) 2668.
  %%CITATION = PHRVA,D16,2668;%%

%\cite{Hassan:2011ea}
\bibitem{Hassan:2011ea}
  S.~F.~Hassan and R.~A.~Rosen,
  ``Confirmation of the Secondary Constraint and Absence of Ghost in Massive Gravity and Bimetric Gravity,''
  arXiv:1111.2070 [hep-th].
  %%CITATION = ARXIV:1111.2070;%%

%\cite{Hassan:2011zd}
\bibitem{Hassan:2011zd}
  S.~F.~Hassan and R.~A.~Rosen,
  ``Bimetric Gravity from Ghost-free Massive Gravity,''
  JHEP {\bf 1202} (2012) 126
  [arXiv:1109.3515 [hep-th]].
  %%CITATION = ARXIV:1109.3515;%%
  
  %\cite{Hassan:2011tf}
\bibitem{Hassan:2011tf}
  S.~F.~Hassan, R.~A.~Rosen and A.~Schmidt-May,
  ``Ghost-free Massive Gravity with a General Reference Metric,''
  JHEP {\bf 1202} (2012) 026
  [arXiv:1109.3230 [hep-th]].
  %%CITATION = ARXIV:1109.3230;%%
  
  
  %\cite{Burrage:2011cr}
\bibitem{Burrage:2011cr}
  C.~Burrage, C.~de Rham, L.~Heisenberg and A.~J.~Tolley,
  ``Chronology Protection in Galileon Models and Massive Gravity,''
  arXiv:1111.5549 [hep-th].
  %%CITATION = ARXIV:1111.5549;%%

%\cite{D'Amico:2011jj}
\bibitem{D'Amico:2011jj}
  G.~D'Amico, C.~de Rham, S.~Dubovsky, G.~Gabadadze, D.~Pirtskhalava and A.~J.~Tolley,
  ``Massive Cosmologies,''
  Phys.\ Rev.\ D {\bf 84} (2011) 124046
  [arXiv:1108.5231 [hep-th]].
  %%CITATION = ARXIV:1108.5231;%%

%\cite{deRham:2011qq}
\bibitem{deRham:2011qq}
  C.~de Rham, G.~Gabadadze and A.~J.~Tolley,
  ``Helicity Decomposition of Ghost-free Massive Gravity,''
  JHEP {\bf 1111} (2011) 093
  [arXiv:1108.4521 [hep-th]].
  %%CITATION = ARXIV:1108.4521;%%


%\cite{deRham:2011rn}
\bibitem{deRham:2011rn}
  C.~de Rham, G.~Gabadadze and A.~Tolley,
  ``Ghost free Massive Gravity in the St\'uckelberg language,''
  arXiv:1107.3820 [hep-th].
  %%CITATION = ARXIV:1107.3820;%%

%\cite{deRham:2011by}
\bibitem{deRham:2011by}
  C.~de Rham and L.~Heisenberg,
  ``Cosmology of the Galileon from Massive Gravity,''
  Phys.\ Rev.\ D {\bf 84} (2011) 043503
  [arXiv:1106.3312 [hep-th]].
  %%CITATION = ARXIV:1106.3312;%%

%\cite{deRham:2010kj}
\bibitem{deRham:2010kj}
  C.~de Rham, G.~Gabadadze and A.~J.~Tolley,
  ``Resummation of Massive Gravity,''
  Phys.\ Rev.\ Lett.\  {\bf 106} (2011) 231101
  [arXiv:1011.1232 [hep-th]].
  %%CITATION = ARXIV:1011.1232;%%

%\cite{deRham:2010ik}
\bibitem{deRham:2010ik}
  C.~de Rham and G.~Gabadadze,
  ``Generalization of the Fierz-Pauli Action,''
  Phys.\ Rev.\ D {\bf 82} (2010) 044020
  [arXiv:1007.0443 [hep-th]].
  %%CITATION = ARXIV:1007.0443;%%

%\cite{deRham:2010gu}
\bibitem{deRham:2010gu}
  C.~de Rham and G.~Gabadadze,
  ``Selftuned Massive Spin-2,''
  Phys.\ Lett.\ B {\bf 693} (2010) 334
  [arXiv:1006.4367 [hep-th]].
  %%CITATION = ARXIV:1006.4367;%%


\bibitem{Crisostomi:2012db}
  M.~Crisostomi, D.~Comelli and L.~Pilo,
  ``Perturbations in Massive Gravity Cosmology,''
  arXiv:1202.1986 [hep-th].
  %%CITATION = ARXIV:1202.1986;%%

\bibitem{Khosravi:2012rk}
  N.~Khosravi, H.~R.~Sepangi and S.~Shahidi,
  ``On massive cosmological scalar perturbations,''
  arXiv:1202.2767 [gr-qc].
  %%CITATION = ARXIV:1202.2767;%%

\bibitem{Volkov:2012wp}
  M.~S.~Volkov,
  ``Hairy black holes in the bigravity theory,''
  arXiv:1202.6682 [hep-th].
  %%CITATION = ARXIV:1202.6682;%%
  
\bibitem{Baccetti:2012bk}
  V.~Baccetti, P.~Martin-Moruno and M.~Visser,
  ``Massive gravity from bimetric gravity,''
  arXiv:1205.2158 [gr-qc].
  %%CITATION = ARXIV:1205.2158;%%
  
\bibitem{Paulos:2012xe}
  M.~F.~Paulos and A.~J.~Tolley,
  ``Massive Gravity Theories and limits of Ghost-free Bigravity models,''
  arXiv:1203.4268 [hep-th].
  %%CITATION = ARXIV:1203.4268;%%
  
\bibitem{DeFelice:2012mx}
  A.~De Felice, A.~E.~Gumrukcuoglu and S.~Mukohyama,
  ``Massive gravity: nonlinear instability of the homogeneous and isotropic universe,''
  arXiv:1206.2080 [hep-th].
  %%CITATION = ARXIV:1206.2080;%%
  
\bibitem{Berezhiani:2011mt}
  L.~Berezhiani, G.~Chkareuli, C.~de Rham, G.~Gabadadze and A.~J.~Tolley,
  ``On Black Holes in Massive Gravity,''
  Phys.\ Rev.\ D {\bf 85} (2012) 044024
  [arXiv:1111.3613 [hep-th]].
  %%CITATION = ARXIV:1111.3613;%%
  
\bibitem{Bergshoeff:2009hq}
  E.~A.~Bergshoeff, O.~Hohm and P.~K.~Townsend,
  ``Massive Gravity in Three Dimensions,''
  Phys.\ Rev.\ Lett.\  {\bf 102} (2009) 201301
  [arXiv:0901.1766 [hep-th]].
  %%CITATION = ARXIV:0901.1766;%%
 
   \bibitem{deRham:2011ca}
  C.~de Rham, G.~Gabadadze, D.~Pirtskhalava, A.~J.~Tolley and I.~Yavin,
  ``Nonlinear Dynamics of 3D Massive Gravity,''
  JHEP {\bf 1106} (2011) 028
  [arXiv:1103.1351 [hep-th]].
  %%CITATION = ARXIV:1103.1351;%%

\bibitem{Afshar:2009rg}
  H.~R.~Afshar, M.~Alishahiha and A.~Naseh,
  ``On three dimensional bigravity,''
  Phys.\ Rev.\ D {\bf 81} (2010) 044029
  [arXiv:0910.4350 [hep-th]].
  %%CITATION = ARXIV:0910.4350;%%

\bibitem{Hinterbichler:2012cn}
  K.~Hinterbichler and R.~A.~Rosen,
  ``Interacting Spin-2 Fields,''
  arXiv:1203.5783 [hep-th].
  %%CITATION = ARXIV:1203.5783;%%

\bibitem{Hassan:2012wt}
  S.~F.~Hassan, A.~Schmidt-May and M.~von Strauss,
  ``Metric Formulation of Ghost-Free Multivielbein Theory,''
  arXiv:1204.5202 [hep-th].
  %%CITATION = ARXIV:1204.5202;%%
  
\bibitem{Khosravi:2011zi}
  N.~Khosravi, N.~Rahmanpour, H.~R.~Sepangi and S.~Shahidi,
  ``Multi-Metric Gravity via Massive Gravity,''
  Phys.\ Rev.\ D {\bf 85} (2012) 024049
  [arXiv:1111.5346 [hep-th]].
  %%CITATION = ARXIV:1111.5346;%%

\bibitem{ArkaniHamed:2002sp}
  N.~Arkani-Hamed, H.~Georgi and M.~D.~Schwartz,
  %``Effective field theory for massive gravitons and gravity in theory space,''
  Annals Phys.\  {\bf 305} (2003) 96
  [hep-th/0210184].
  %%CITATION = HEP-TH/0210184;%%
  
\bibitem{Creminelli:2005qk}
  P.~Creminelli, A.~Nicolis, M.~Papucci and E.~Trincherini,
  ``Ghosts in massive gravity,''
  JHEP {\bf 0509} (2005) 003
  [hep-th/0505147].
  %%CITATION = HEP-TH/0505147;%%

%\cite{vanDam:1970vg}
\bibitem{vanDam:1970vg}
  H.~van Dam and M.~J.~G.~Veltman,
  ``Massive and massless Yang-Mills and gravitational fields,''
  Nucl.\ Phys.\ B {\bf 22} (1970) 397.
  %%CITATION = NUPHA,B22,397;%%
  
  %\cite{Zakharov:1970cc}
\bibitem{Zakharov:1970cc}
  V.~I.~Zakharov,
  ``Linearized gravitation theory and the graviton mass,''
  JETP Lett.\  {\bf 12} (1970) 312
   [Pisma Zh.\ Eksp.\ Teor.\ Fiz.\  {\bf 12} (1970) 447].
  %%CITATION = JTPLA,12,312;%%
  
\bibitem{Vainshtein:1972sx}
  A.~I.~Vainshtein,
  ``To the problem of nonvanishing gravitation mass,''
  Phys.\ Lett.\ B {\bf 39} (1972) 393.
  %%CITATION = PHLTA,B39,393;%%

\bibitem{Weinberg:2008zzc}
  S.~Weinberg,
  ``Cosmology,''
  Oxford, UK: Oxford Univ. Pr. (2008) 593 p

\bibitem{Alberte:2011ah}
  L.~Alberte,
  ``Massive Gravity on Curved Background,''
  arXiv:1110.3818 [hep-th].
  %%CITATION = ARXIV:1110.3818;%%
  
\bibitem{Higuchi:1986py}
  A.~Higuchi,
  ``Forbidden Mass Range For Spin-2 Field Theory In De Sitter Space-time,''
  Nucl.\ Phys.\ B {\bf 282} (1987) 397.
  %%CITATION = NUPHA,B282,397;%%

  
\bibitem{Deser:2001wx}
  S.~Deser and A.~Waldron,
   ``Stability of massive cosmological gravitons,''
  Phys.\ Lett.\ B {\bf 508} (2001) 347
  [hep-th/0103255].
  %%CITATION = HEP-TH/0103255;%%


\end{thebibliography}
\end{document}